\documentclass[longauth]{aa}  

\usepackage{amsmath}
\usepackage{booktabs}
\usepackage{enumitem}
\usepackage{flushend}
\usepackage{graphicx}
\usepackage{ifthen}
\usepackage{lastpage}
\usepackage{siunitx}
\usepackage{txfonts}
\usepackage{xspace}

\graphicspath{{fig/}}

\newcommand{\uncertainty}[3]{#1\,^{+\,#2}_{-\,#3}}
\newcommand\galaxyname[2]{#1\,#2}
\newcommand\ionized[2]{[\mathrm{#1}\,\textsc{#2}]}
\newcommand\OI[1][]{\ifthenelse{\equal{#1}{}}{\ionized{O}{i}}{\ionized{O}{i}\,\lambda#1}}
\newcommand\OIII[1][]{\ifthenelse{\equal{#1}{}}{\ionized{O}{iii}}{\ionized{O}{iii}\,\lambda#1}}
\newcommand\NII[1][]{\ifthenelse{\equal{#1}{}}{\ionized{N}{ii}}{\ionized{N}{ii}\,\lambda#1}}
\newcommand\SII[1][]{\ifthenelse{\equal{#1}{}}{\ionized{S}{ii}}{\ionized{S}{ii}\,\lambda#1}}
\newcommand\SIII[1][]{\ifthenelse{\equal{#1}{}}{\ionized{S}{iii}}{\ionized{S}{iii}\,\lambda#1}}
\newcommand\HA[1][]{\ifthenelse{\equal{#1}{}}{\mathrm{H}\,\alpha}{\mathrm{H}\,\alpha\,\lambda#1}}
\newcommand\HB[1][]{\ifthenelse{\equal{#1}{}}{\mathrm{H}\,\beta}{\mathrm{H}\,\beta\,\lambda#1}}
\newcommand\HII{H\,\textsc{ii}\xspace}
\newcommand\fesc[1][]{\ifthenelse{\equal{#1}{}}{f_\mathrm{esc}}{f_\mathrm{esc,#1}}}
\newcommand\DeltaOH{\Delta (\mathrm{O}/\mathrm{H})}
\newcommand\Qpred{Q(\mathrm{H}^0)}
\newcommand\Qobs{Q_{\HA}}

\DeclareSIUnit\percent{per\,cent}
\DeclareSIUnit\arcsec{arcsec}
\DeclareSIUnit\arcmin{arcmin}
\DeclareSIUnit\parsec{pc}
\DeclareSIUnit\year{yr}
\DeclareSIUnit\erg{erg}
\DeclareSIUnit\Msun{M_\odot}
\DeclareSIUnit\angstrom{\text{Å}}

\defcitealias{Scheuermann+2023}{Paper~I}

\usepackage[
    colorlinks=true,
    linkcolor=blue,
    citecolor=blue,
    filecolor=blue,
    urlcolor=blue
    ]{hyperref}
\usepackage[all]{hypcap} 
\usepackage[
    noabbrev,
    nameinlink,
    capitalise
    ]{cleveref} 	
\usepackage{orcidlink}

\begin{document}

\title{Stellar associations powering \HII regions}
\subtitle{II. Escape fraction of ionizing photons}
\titlerunning{Escape fraction of ionizing photons}
\author{
    Fabian Scheuermann\thanks{f.scheuermann@uni-heidelberg.de}\inst{\ref{inst:ARI}}\orcidlink{0000-0003-2707-4678}\and
    Kathryn Kreckel\inst{\ref{inst:ARI}}\orcidlink{0000-0001-6551-3091}\and
    Jia Wei Teh\inst{\ref{inst:ITA}}\orcidlink{0000-0001-7863-5047}\and 
    Francesco Belfiore\inst{\ref{inst:ESO},\ref{inst:INAF}}\orcidlink{0000-0002-2545-5752}\and
    Brent Groves\inst{\ref{inst:Perth},\ref{inst:AAL}}\orcidlink{0000-0002-9768-0246}\and
    Ashley T.\ Barnes \inst{\ref{inst:ESO}}\orcidlink{0000-0003-0410-4504} \and
    Médéric Boquien\inst{\ref{inst:UniCA}}\orcidlink{0000-0003-0946-6176} \and
    M\'elanie Chevance\inst{\ref{inst:ITA}}\orcidlink{0000-0002-5635-5180}\and
    Daniel A.\ Dale\inst{\ref{inst:Wyoming}}\orcidlink{0000-0002-5782-9093}\and
    Oleg Egorov\inst{\ref{inst:ARI}}\orcidlink{0000-0002-4755-118X}\and
    Simon C.\ O.\ Glover\inst{\ref{inst:ITA}}\orcidlink{0000-0001-6708-1317}\and 
    Kathryn Grasha\inst{\ref{inst:Canberra}}\orcidlink{0000-0002-3247-5321}\and
    Stephen Hannon\inst{\ref{inst:MPIA}}\orcidlink{0000-0001-9628-8958}\and
    Ralf S.\ Klessen\inst{\ref{inst:ITA},\ref{inst:IWR}}\orcidlink{0000-0002-0560-3172}\and
    Kirsten L.\ Larson\inst{\ref{inst:ESA_STsci}}\orcidlink{0000-0003-3917-6460}\and
    Janice C.\ Lee\inst{\ref{inst:STsci}}\orcidlink{0000-0002-2278-9407}\and
    Fu-Heng Liang\inst{\ref{inst:ARI}}\orcidlink{0000-0003-2496-1247}\and
    Laura A.\ Lopez\inst{\ref{inst:Ohio}}\orcidlink{0000-0002-1790-3148}\and
    J.\ Eduardo M\'endez-Delgado\inst{\ref{inst:Mexico}}\orcidlink{0000-0002-6972-6411}\and 
    Justus Neumann\inst{\ref{inst:MPIA}}\orcidlink{0000-0002-3289-8914}\and 
    Eve Ostriker\inst{\ref{inst:Princeton}}\orcidlink{0000-0002-0509-9113}\and
    Hsi-An Pan\inst{\ref{inst:Taiwan}}\orcidlink{0000-0002-1370-6964}\and
    Lise Ramambason\inst{\ref{inst:ITA}}\orcidlink{0000-0002-9190-9986}\and
    Francesco Santoro\inst{\ref{inst:MPIA}}\orcidlink{0000-0002-6363-9851}\and
    Eva Schinnerer\inst{\ref{inst:MPIA}}\orcidlink{0000-0002-3933-7677}\and
    David A.\ Thilker\inst{\ref{inst:Hopkins}}\orcidlink{0000-0002-8528-7340}\and
    Qiushi Chris Tian\inst{\ref{inst:Hopkins},\ref{inst:Leiden}}\orcidlink{0009-0009-9148-2159}\and
    Leonardo \'Ubeda\inst{\ref{inst:STsci}}\orcidlink{0000-0001-7130-2880}\and
    Bradley C.\ Whitmore\inst{\ref{inst:STsci}}\orcidlink{0000-0002-3784-7032}\and
    Thomas G.\ Williams\inst{\ref{inst:JBCA}}\orcidlink{0000-0002-0012-2142}
}
\authorrunning{Scheuermann et al.}
\institute{
    Astronomisches Rechen-Institut, Zentrum f\"{u}r Astronomie der Universit\"{a}t Heidelberg, M\"{o}nchhofstra\ss e 12-14, 69120 Heidelberg, Germany\label{inst:ARI}\and
    Universit\"{a}t Heidelberg, Zentrum f\"{u}r Astronomie, Institut f\"{u}r Theoretische Astrophysik, Albert-Ueberle-Str.\ 2, 69120, Heidelberg, Germany\label{inst:ITA}\and 
    European Southern Observatory (ESO), Karl-Schwarzschild-Stra{\ss}e 2, 85748 Garching, Germany\label{inst:ESO}\and
    INAF -- Osservatorio Astrofisico di Arcetri, Largo E.\ Fermi 5, I-50157 Firenze, Italy\label{inst:INAF}\and 
    International Centre for Radio Astronomy Research, University of Western Australia, 7 Fairway, Crawley, 6009 WA, Australia\label{inst:Perth}\and
    Astronomy Australia Ltd.\ (Perth Office), University of Western Australia, 7 Fairway, Crawley, 6009 WA, Australia\label{inst:AAL}\and
    Université Côte d'Azur, Observatoire de la Côte d'Azur, CNRS, Laboratoire Lagrange, 06000 Nice, France\label{inst:UniCA}\and
    Department of Physics \& Astronomy, University of Wyoming, Laramie, WY 82071, USA\label{inst:Wyoming}\and
    Research School of Astronomy and Astrophysics, Australian National University, Canberra, ACT 2611, Australia\label{inst:Canberra}\and
    Max Planck Institut für Astronomie, K\"{o}nigstuhl 17, 69117 Heidelberg, Germany\label{inst:MPIA}\and 
    Universit\"{a}t Heidelberg, Interdisziplin\"{a}res Zentrum f\"{u}r Wissenschaftliches Rechnen, Im Neuenheimer Feld 225, 69120 Heidelberg, Germany\label{inst:IWR}\and
    AURA for the European Space Agency (ESA), Space Telescope Science Institute, 3700 San Martin Drive, Baltimore, MD 21218, USA\label{inst:ESA_STsci}\and 
    Space Telescope Science Institute, 3700 San Martin Dr., Baltimore, MD 21218, USA\label{inst:STsci}\and 
    Department of Astronomy, The Ohio State University, Columbus, OH 43210, USA\label{inst:Ohio}\and
    Instituto de Astronomía, Universidad Nacional Autónoma de México, Ap.\ 70-264, 04510 CDMX, Mexico\label{inst:Mexico}\and
    Department of Astrophysical Sciences, Princeton University, 4 Ivy Lane, Princeton, NJ 08544, USA\label{inst:Princeton}\and
    Department of Physics, Tamkang University, No.151, Yingzhuan Road, Tamsui Dist., New Taipei City 251301, Taiwan\label{inst:Taiwan}\and
    Department of Physics and Astronomy, The Johns Hopkins University, Baltimore, MD 21218, USA\label{inst:Hopkins}\and
    Leiden Observatory, Leiden University, P.O.\ Box 9513, 2300 RA Leiden, The Netherlands\label{inst:Leiden}\and
    UK ALMA Regional Centre Node, Jodrell Bank Centre for Astrophysics, Department of Physics and Astronomy, The University of Manchester, Oxford Road, Manchester M13 9PL, UK\label{inst:JBCA}
} 
\date{Received 10 August 2025 / Accepted 7 February 2026}
\abstract{
    Newly formed stars have a profound impact on their environment by depositing energy and momentum into the surrounding gas. 
    However, only a fraction of the stellar feedback is retained in the cloud and observational constraints are needed to further our understanding of this process. 
    In a sample of 19 nearby galaxies, we match \HII regions from PHANGS--MUSE to their ionizing stellar source from PHANGS--\textit{HST} and measure the percentage of ionizing radiation that is leaking into the surrounding diffuse ionized gas (DIG). 
    Based on a catalogue, where each \HII region is powered by a single young and massive stellar association, we measure a photon escape fraction of $\fesc=82^{+12}_{-24}\;\si{\percent}$. 
    Comparable results are obtained when different procedures are used to match the ionized gas to its source. 
    All samples we study contain a substantial fraction of objects (up to \SI{20}{\percent}), where the stellar source is not sufficient to produce the $\HA$ flux observed from the nebula. 
    Many of them are probably related to uncertain age estimates, but we also find numerous regions, where a significant fraction of the ionizing photon budget is contributed by stars that reside outside the boundaries of the \HII region. 
    This motivates the use of an alternative galaxy-wide approach, in which we include all \HII regions and stellar sources, not just the ones that show a clear overlap. 
    When summing up the ionization budget over entire galaxies, we measure slightly lower, but consistent values. 
}
\keywords{
    ISM: \HII regions -- 
    Galaxies: ISM -- 
    Galaxies: star clusters: general --  
    Galaxies: star formation
}
\maketitle
\nolinenumbers

\section{Introduction}\label{sec:introduction}

Stellar feedback plays a crucial role in the evolution of the interstellar medium \citep[ISM,][]{Lopez+2014,Klessen+2016,Krumholz+2019}. 
It shapes the surrounding gas, halts or triggers new star formation, and enriches the gas with metals. 
However, not all energy and momentum from the stellar feedback is deposited in the \HII regions \citep{Hanish+2010}. 
Turbulence and radiative feedback can fragment the cloud \citep{Kakiichi+2021}, which opens up channels through which some radiation escapes, while part of the stellar feedback is absorbed by gas and dust. 
This is an important contribution to the diffuse ionized gas \citep[DIG,][]{Haffner+2009}. 

A common way to constrain the leakage is by comparing the number of predicted ionizing photons emitted by individual stars or clusters, $\Qpred$, to the ionizing photon flux inferred from the observed $\HA$ luminosity of an associated \HII region, $\Qobs$ \citep{Niederhofer+2016}. 
This yields the escape fraction of ionizing photons which is usually defined as 
\begin{equation}\label{eq:escape_fraction}
    \fesc = \frac{\Qpred - \Qobs}{\Qpred} . 
\end{equation}
It is expected to evolve over time \citep{Trebitsch+2017} and depend on various local ISM conditions, such as gas density, metallicity and the star-gas geometry \citep{Ramambason+2022}. 
Therefore, having a large sample is necessary to marginalize over these variations, or explore trends. 

For \HII regions located within a massive galaxy, any escaping radiation is expected to primarily impact the surrounding galactic ISM and it is clear that within galaxies there are $\si{\kilo\parsec}$-scale consequences of stellar feedback \citep{Zurita+2000, Belfiore+2022}. 
Connecting this feedback to the galactic escape fraction, which is vital to understanding cosmic reionization \citep{Paardekooper+2011, Mitra+2013,Japelj+2017,KadoFong+2020,Ma+2020,Ramambason+2020,Wang+2021,Chisholm+2022,SaldanaLopez+2022,Flury+2022a,Flury+2022b}, is challenging. 

Measurements of $\fesc$ with various techniques have resulted in a wide range of estimates. 
One of the first attempts that applied a method similar to ours, i.e.\ analysing individual ionizing sources in nearby galaxies, was conducted by \citet{McLeod+2020} for \galaxyname{NGC}{300}. 
Because the galaxy is nearby (\SI{2.09}{\mega\parsec}; \citealt{Jacobs+2009}), they were able to use the Multi Unit Spectroscopic Explorer (MUSE) to identify individual O-stars as the source of the ionizing radiation. 
By matching them to two \HII regions they measured an escape fraction of \SI{28 \pm 6}{\percent} and \SI{51 \pm 1}{\percent} respectively. 
\citet{DellaBruna+2021} analysed \num{8} \HII regions in \galaxyname{NGC}{7793} and found a value of $\fesc=67^{+\ 8}_{-12}\;\si{\percent}$. 
In \citet{DellaBruna+2022b}, they applied their technique to \galaxyname{M}{83} and found \num{541} matches between \HII regions and young star clusters. 
But only \SI{13}{\percent} of them returned a positive $\fesc$, i.e.\ for most objects, the photons emitted by the stars were insufficient to explain the observed $\HA$ emission. 
Recently, \citet{Teh+2023} studied \num{42} \HII regions in \galaxyname{NGC}{628} and determined an escape fraction of $\uncertainty{9}{6}{6}\;\si{\percent}$. 

Studies of even closer galaxies do not give more consistent results, although individual ionizing stars can be characterised and rich multi-wavelength constraints are available. 
\citet{Pellegrini+2012} analysed \num{13} \HII regions in the Large Magellanic Cloud with known ionizing stars and infer typical escape fractions ranging from \SIrange{30}{60}{\percent}. 
However, in the extreme starburst region of the Tarantula Nebula, \citet{Doran+2013} measured extremely low escape fractions of \SIrange{5}{10}{\percent}. 
The dwarf galaxies Holmberg I \citep{Egorov+2018} and Sextans A \citep{Gerasimov+2022} have well characterised O star populations and escape fractions of $\sim\SI{60}{\percent}$, but larger dwarf galaxy samples, where modelling included infrared constraints, typically return much lower \SIrange{10}{20}{\percent} escape fractions. 

Within simulations, considerable progress has been made in recent years to include more realistic treatments of stellar feedback physics, but again a range of escape fractions are predicted as a function of both local conditions and time. 
Cloud-scale simulations with simplified feedback prescriptions \citep[e.g.][]{Dale+2012,Dale+2013} indicate that the escape fraction can vary widely over time as photoionization reshapes the cloud structure. 
Radiation-hydrodynamic models of turbulent clouds, some including magnetic fields \citep{Howard+2017,Howard+2018,Kim+2019,Kim+2021a}, also find strongly time-dependent escape fractions that can briefly approach $\SI{100} {\percent}$ but average to only a few to ten \si{\percent} over a cloud’s lifetime. 
Semi-analytic feedback models \citep{Rahner+2017} predict similar trends, with escape fractions rising as feedback clears gas, while galaxy-scale simulations \citep{Tacchella+2022} infer global values of order a few \si{\percent}. 

With the availability of large field-of-view integral field unit (IFU) spectroscopy instruments, like MUSE or the Spectromètre Imageur à Transformée de Fourier pour l'Etude en Long et en Large de raies d'Emission (SITELLE), it is now possible to study the ionized gas using multiple optical emission line diagnostics, isolating individual \HII regions while imaging the full star-forming disk for a large sample of nearby galaxies. 
In combination with space-based imaging (e.g.\ \textit{Hubble Space Telescope; HST}) we are able to obtain an accurate census of the ionizing sources in galaxies beyond the Local Group. 
With its combination of MUSE and \textit{HST} imaging, the Physics at High Angular resolution in Nearby GalaxieS (PHANGS)\footnote{\url{https://www.phangs.org}} survey is ideally suited to the systematic study of star formation in nearby galaxies \citep{Leroy+2021a,Leroy+2021b,Emsellem+2022,Lee+2022,Lee+2023}. 
In this work, we use the large sample of \HII regions and stellar associations compiled in \citet{Scheuermann+2023} to calculate $\fesc$ and search for systematic variations with local physical conditions and star cluster properties. 

This paper is organized as follows: in \cref{sec:data} we present the catalogues and models that were used in this work. 
In \cref{sec:escape} we use these data to compute the escape fractions of the \HII regions and discuss trends with physical properties. 
We conclude in \cref{sec:conclusion_escape}.

\section{Data and models}\label{sec:data}

Our sample consists of 19 PHANGS galaxies that were observed with both MUSE and \textit{HST} and is listed in \cref{tbl:sample}. 
The matching of the nebulae and ionizing sources is described in detail in \citet{Scheuermann+2023}, hereafter referred to as \citetalias{Scheuermann+2023}. 
Below we provide a short summary of the individual data sets and how they are combined to form matched catalogues, as well as the stellar models that are used during the analysis. 

\begin{table}
    \centering
    \caption[Galaxy sample for the escape fractions with number the of \HII regions, stellar associations and clusters.]{Properties of the PHANGS galaxy sample.}
    \begin{tabular}{
        l
        S[table-format=2.2]@{\,\( \pm \)\,}
        S[table-format=1.2]
        S[table-format=5]
        S[table-format=5]
        S[table-format=5]
        }
        \toprule
        \multicolumn{1}{c}{Name} & \multicolumn{2}{c}{Distance} & $N_\mathrm{\HII}$ & $N_\mathrm{asc}$ & $N_\mathrm{clu}$ \\
         & \multicolumn{2}{c}{($\si{\mega\parsec}$)} & & & \\
        \midrule
        IC\,5332 & 9.01 & 0.41 & 608 & 397 & 193 \\
        NGC\,0628 & 9.84 & 0.63 & 1651 & 1119 & 1365 \\
        NGC\,1087 & 15.85 & 2.24 & 892 & 487 & 926 \\
        NGC\,1300 & 18.99 & 2.85 & 1147 & 401 & 797 \\
        NGC\,1365 & 19.57 & 0.78 & 445 & 489 & 833 \\
        NGC\,1385 & 17.22 & 2.58 & 922 & 525 & 997 \\
        NGC\,1433 & 18.63 & 1.86 & 729 & 494 & 390 \\
        NGC\,1512 & 18.83 & 1.88 & 479 & 521 & 519 \\
        NGC\,1566 & 17.69 & 2.00 & 1448 & 1582 & 2142 \\
        NGC\,1672 & 19.40 & 2.91 & 1047 & 1247 & 1628 \\
        NGC\,2835 & 12.22 & 0.94 & 777 & 649 & 728 \\
        NGC\,3351 & 9.96 & 0.33 & 769 & 708 & 595 \\
        NGC\,3627 & 11.32 & 0.48 & 1016 & 1325 & 2320 \\
        NGC\,4254 & 13.10 & 2.80 & 2375 & 1661 & 3897 \\
        NGC\,4303 & 16.99 & 3.04 & 1956 & 1736 & 3188 \\
        NGC\,4321 & 15.21 & 0.49 & 984 & 1483 & 1921 \\
        NGC\,4535 & 15.77 & 0.37 & 1168 & 469 & 567 \\
        NGC\,5068 & 5.20 & 0.21 & 1405 & 715 & 1137 \\
        NGC\,7496 & 18.72 & 2.81 & 523 & 265 & 385 \\\midrule
        Total & \multicolumn{2}{c}{} & 20341 & 16273 & 24529 \\
        \bottomrule\addlinespace
    \end{tabular}
    \tablefoot{Distances compiled by \citet{Anand+2021a}. $N_\mathrm{\HII}$ is the number of \HII regions, $N_\mathrm{asc}$ the number of stellar associations, and $N_\mathrm{clu}$ the number of clusters within the overlapping coverage of MUSE and \textit{HST}.}
    \label{tbl:sample}
\end{table}

\subsection{MUSE HII regions}\label{sec:data:HII_regions}

MUSE is a powerful optical IFU spectrograph that is mounted at the Very Large Telescope (VLT) in Chile \citep{Bacon+2010}. 
It possesses a $1\times\SI{1}{\arcmin\squared}$ field of view (FoV) and covers the wavelength range from \SIrange{4800}{9300}{\angstrom} in the nominal mode. 
The PHANGS--MUSE survey (PI: Schinnerer) has produced wide-field mosaics of 19 nearby spiral galaxies. 
The analysis and the main data products are described in \citet{Emsellem+2022}. 
Among these are continuum subtracted emission-line maps, constructed by fitting single Gaussians to a set of bright emission lines (e.g.\ $\HB$, $\OIII\lambda$5007, $\HA$, $\NII\lambda$6583, $\SII\lambda\lambda$6717, 6731, $\SIII\lambda$9531), which form the basis for our further analysis. 

Based on the $\HA$ morphologies, \citet{Santoro+2022} and \citet{Groves+2023} used this dataset to construct an \HII region catalogue. 
In the first step, \mbox{\textsc{hiiphot}} \citep{Thilker+2000} was used to draw the boundaries of the emission line nebulae based on their $\HA$ brightness. 
This initial catalogue is composed of \num{31497} nebulae, but still contains other $\HA$ bright objects like planetary nebulae (PNe) or supernova remnants (SNRs). 
They are removed by analysing diagnostic line ratios in the Baldwin–Phillips–Terlevich (BPT) diagrams \citep{Baldwin+1981,Veilleux+1987}. 
We require regions to fall below the \citet{Kauffmann+2003} diagnostic curve in the $\OIII/\HB$ versus $\NII/\HA$ diagrams, and below the \citet{Kewley+2006} diagnostic curve in the $\OIII/\HB$ versus $\SII/\HA$ diagrams, with $\mathrm{S/N} > 5$ in all lines. 
Where $\OI$ is detected with $\mathrm{S/N} > 5$, we further require that regions fall below the \citet{Kewley+2006} diagnostic curve in the $\OI/\HB$ versus $\SII/\HA$ diagram. 

For this work, we only use the objects that fall inside the \textit{HST} coverage, yielding \num{20341} \HII regions in the final catalogue. 
In addition, properties like metallicity, ionization parameter and density are derived from the measured emission lines. 
For a full summary of the catalogue and the derived properties with detailed statistics, see also \citetalias{Scheuermann+2023}. 

We are primarily interested in the \HII region $\HA$ luminosity, which we want to compare to the predicted ionizing photon rate from the stars. 
The flux measurements for the nebulae are very precise, with a typical uncertainty of less than $\SI{1}{\percent}$, but the uncertainties arising from the determination of the boundaries of the nebulae are not taken into account here. 
We need to correct the measured fluxes for extinction and for the Milky Way, we use the extinction curve from \citet{ODonnell+1994}, $E(B-V)$ from \citet{Schlafly+2011} and $R_V=3.1$. 
For the internal extinction, we adopt the same extinction curve and determine $E(B-V)$ from the Balmer decrement, assuming $\HA/\HB=2.86$. 
To convert the fluxes to luminosities we use the distances listed in \cref{tbl:sample}. 

In order to compare those values with the emission from the stars, we have to convert the $\HA$ luminosity to the number of ionizing photons. 
This is commonly done with a conversion factor like the one by \citet{Hummer+1987}. 
Assuming Case B recombination, an electron temperature of 
$\SI{e4}{\kelvin}$ and a density of $\SI{e2}{\per\cubic\cm}$ (well matched to our sample; \citealt{Barnes+2021}) this yields
\begin{equation}
    \Qobs\,/\, \si{\per \s}= \num{7.31e11} L(\HA)\,/\, \si{\erg\per\s} . 
\end{equation}
It should be noted that the temperatures observed in our \HII regions can vary between \SIrange[range-phrase = { and }]{5000}{10000}{\kelvin} \citep{Kreckel+2022}. 
We therefore run \textsc{intrat} by \citet{Storey+1995} to compute the conversion factor for a number of temperatures and densities. 
A lower temperature of $\SI{5000}{\kelvin}$ would decrease the conversion factor by about $\SI{8}{\percent}$, and imply that we slightly underestimate the number of ionizing photons (and hence overestimate the escape fraction) for some objects. 
An increase of the density to $\SI{e3}{\per\cubic\cm}$ does not change the result by more than $\SI{0.3}{\percent}$.

\subsection{\textit{HST} stellar associations and star clusters}\label{sec:data:stars}

All 19 galaxies are included in the PHANGS--\textit{HST} survey \citep{Lee+2022}, providing a sharp $\sim$\SI{4}{pc} view of ionizing stellar sources with coverage in five broadband filters: F275W (NUV), F336W (\textit{U}), F438W (\textit{B}), F555W (\textit{V}), and F814 (\textit{I}). 
From these images, we can identify and distinguish both compact, gravitationally bound star clusters, as well as physically larger, loosely bound stellar associations \citep{Elmegreen+2008,Gouliermis+2018}. 
In \citetalias{Scheuermann+2023} we found that stellar associations provide a better one-to-one match to individual \HII regions, however compact star clusters are easier to identify and model, and have been the focus of most previous studies constraining \HII region escape fractions \citep[e.g.][]{DellaBruna+2021,Teh+2023}. 
Therefore, in this work we include both star clusters and stellar associations in our analysis, flexibly allowing \HII regions to contain multiple ionizing sources. 

The stellar association catalogue was created by \citet{Larson+2023}, and employs a watershed algorithm to join nearby young stars into a single object. 
Starting from either NUV or \textit{V} tracer images, \textsc{dolphot} is used to identify an initial set of point sources. 
This is then used to construct density maps, which are Gaussian smoothed to physical scales ranging from \SIrange{8}{64}{pc}. 
A watershed algorithm finally groups multiple peaks together in order to form a single larger structure, or stellar association. 
Using the integrated fluxes of all stellar peaks within the association provides five band photometry of each catalogued object. 
As demonstrated in \citetalias{Scheuermann+2023}, our analysis yields similar results when we use different scales and tracer bands, and we decide to use the \SI{32}{\parsec} NUV catalogue for our analysis, as it has the largest overlap with our \HII region catalogue. 
This catalogue contains \num{16273} associations across 19 galaxies that fall inside the MUSE coverage. 
This number differs slightly from the one used in \citetalias{Scheuermann+2023}. 
The previous association catalogue was an internal release and some associations in \galaxyname{NGC}{0628} lay outside the overlapping footprint of all five \textit{HST} bands. 
This caused some issues with the SED fit and hence those few objects are now removed from the catalogue. 

For the star clusters we use the catalogue created by \citet{Thilker+2022} and \citet{Maschmann+2024}, which identifies marginally resolved sources in \textit{HST} imaging and characterises them based on their morphology. 
The sources are detected with \textsc{dolphot} and \textsc{daostarfinder} \citep[v2.0,][]{Dolphin+2000,Stetson+1987} and classified into Classes 1, 2, 3, and 4 using machine learning models \citep{Wei+2020,Hannon+2023}. 
Class 1 and 2 objects represent the compact clusters while Class 3 objects are often multi-peaked and are typically referred to as ``compact associations'' of stars. 
Class 4 objects are not considered to be star clusters, but instead are classified as artefacts like background galaxies or single stars, and are therefore excluded. 
As with the stellar associations, integrated fluxes are obtained for each object to provide five band photometry. 
The catalogue that we use contains \num{24529} star clusters that fall inside the MUSE coverage (\num{6593} of Class 1 or 2 and \num{17936} of Class 3). 

For both the stellar association and compact star cluster catalogues, the age, mass, and reddening were derived by fitting the \textit{HST} five band spectral energy distribution (SED) as described in \citet{Turner+2021}. 
To do this, Code Investigating GALaxy Emission (\textsc{cigale}) by \citet{Boquien+2019} was used with theoretical models of a single stellar population by \citet{Bruzual+2003}, a metallicity of $Z=0.02$, and assuming a fully-sampled \citet{Chabrier+2003} initial mass function (IMF). 
Since the publication of \citetalias{Scheuermann+2023}, the cluster catalogue was updated by \citet{Thilker+2025}. 
However, they are broadly consistent and the changes do not alter our results, so for reproducibility we carry out our analysis on the catalogue presented in \citetalias{Scheuermann+2023}. 

We note a few caveats on the use of these catalogues to study the youngest stellar populations. 
A limited colour-space is used to derive the stellar properties, with five \textit{HST} bands used to fit three stellar properties (mass, age, and reddening). 
While this will be refined in the near future with the inclusion of additional \textit{JWST} bands \citep{Lee+2023}, the lack of suitable FUV coverage still presents a problem, especially for the youngest clusters. 
While the galaxies were observed with \textit{GALEX} \citep{Martin+2005} and \textit{AstroSat} \citep{Hassani+2024}, the resolution of those instruments are not sufficient for the analysis. 
Another issue that we cannot address with our current data is the contribution of Wolf-Rayet stars \citep{Crowther+2007} or other massive single stars \citep{Zastrow+2013} to the ionizing photon output.

\subsection{Calculating the number of ionizing photons}\label{sec:population_synthesis}

What is missing in the published catalogues is the number of ionizing photons $\Qpred$ that is emitted by the stars. 
This number can depend on a variety of assumptions, such as the adopted stellar atmosphere model \citep{SimonDiaz+2008}, stellar rotation \citep{Levesque+2012}, and binarity \citep{Lecroq+2024}. 
In this section we use a number of different codes to compute $\Qpred$ as a way of understanding the systematics that go into this calculation. 

Generally, the models take a certain IMF to create a grid of stars and then compute the ionizing photon flux as a function of age, based on different stellar models and atmospheres. 
If the IMF is fully sampled, the output scales linearly with mass. 
This should be treated with caution, as this is certainly not the case for lower mass clusters \citep{Cervino+2006}. 
We try to circumvent this problem by applying certain mass cuts, as discussed in \cref{sec:matched_catalogues}. 
To propagate the uncertainties from the ages, we assume a Gaussian age distribution that is clipped at $\SI{1}{\mega\year}$ and draw a random sample for which we then compute $\Qpred$. 
Since the flux is expected to scale linearly with mass (for a fully-sampled IMF), error propagation is straightforward. 
The uncertainty of the distance affects the $\HA$ luminosity and the derived cluster mass in the same way and is therefore not included. 

For the comparison, we consider results from \textsc{cigale}, \textsc{starburst99} (the models without rotation \texttt{SB99 v00} and those with rotation \texttt{SB99 v40}), and the binarity model from the Binary Population and Spectral Synthesis code (\textsc{bpass}). 
An overview of the details and the age-dependency of different models is shown in \cref{sec:appendix_models}. 
Another popular model is Stochastically Lighting Up Galaxies \citep[\textsc{slug},][]{daSilva+2012}, but instead of discrete values, it treats the individual properties as probability distributions. 
This makes it difficult to include here and we refer to \cref{sec:appendix_slug} for some simplistic comparisons of our results with what is modelled by \textsc{slug}.

\subsubsection{{\upshape\texttt{BC03}}}

As previously stated, \textsc{cigale} \citep{Boquien+2019} was used to fit the SED and derive stellar properties. 
Therefore the self-consistent approach is to also use the ionizing photon flux that is computed by this code. 
The catalogues published by \citet{Thilker+2022} and \citet{Larson+2023} do not contain this quantity, so we run \textsc{cigale} with the same input parameters in \texttt{savefluxes} mode to compute the ionizing photon flux as a function of age. 
We then use the previously fitted ages and masses of each stellar cluster and association to interpolate their value of $\Qpred$. 
The stellar models from \citet{Bruzual+2003} are used, together with an IMF from \citet{Chabrier+2003}.

\subsubsection{{\upshape\texttt{SB99}}}

Another popular stellar population synthesis model is \textsc{starburst99} \citep{Leitherer+1999,Leitherer+2010,Leitherer+2014,Vazquez+2005}. 
As input parameters, we select a single burst of star formation that follows a fully-sampled \citet{Kroupa+2001} IMF with the Geneva stellar models at solar metallicity \citep{Ekstroem+2012} and stellar atmospheres from \citet{Lejeune+1997}. 
The remaining parameters are left at their default values. 
After that we use the stellar ages and masses again to interpolate the output to the values in our sample. 
While it represents more of an extreme case, we also consider a model with rotating stars, spinning at $\SI{40}{\percent}$ break-up velocity, as a limiting case.

\subsubsection{{\upshape\texttt{BPASS}}}

Most models track single star evolution, but many stars form in binaries, which impacts their evolution \citep{Sana+2012}. 
Fortunately for us, the difference in ionizing photons is expected to only become significant after $\SI{8}{\mega\year}$ \citep{Lecroq+2024}, and hence this should not be an issue for our purposes. 
In order to get a rough estimate on the impact binarity might have on $\Qpred$, we include \textsc{bpass} by \citet{Eldridge+2009} in our comparison. 
We employ the fiducial model for binaries from version 2.1 \citep{Eldridge+2017} with a Kroupa IMF \citep{Kroupa+2001} and an upper mass limit of \SI{300}{\Msun} at solar metallicity.

\subsubsection{Comparison between the different models}

\begin{figure}
\centering
\includegraphics[width=\columnwidth]{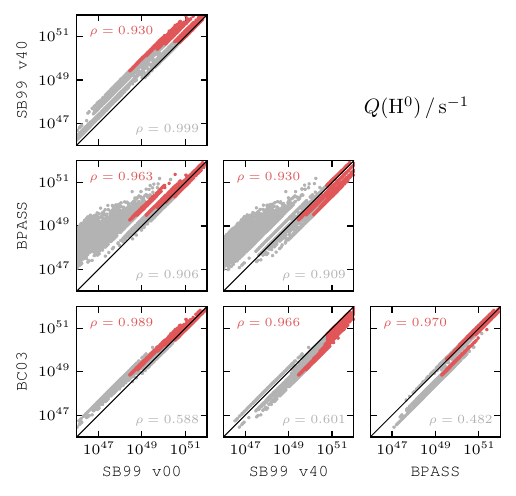}
\caption[Comparison between different population synthesis models.]{Comparison between different population synthesis models. 
We use a sample with random ages and masses and compute the ionizing photon flux $\Qpred$ with \texttt{BC03}, \texttt{SB99}, and \texttt{BPASS}. 
The grey points are the full sample and the red points show a young and massive subsample ($\mathrm{age}\leq\SI{8}{\mega\year}$ and mass $>\SI{e4}{\Msun}$). 
The Spearman correlation coefficient $\rho$ indicates good agreement between the different models across the robust subsamples, and all of the correlations are statistically significant with $p\text{-values}<0.05$.}
\label{fig:corner_ionizing_photons}
\end{figure}

To gauge the variations between the different models introduced in the previous sections, we compare the fluxes they predict for a random sample. 
For this, we create \num{10000} clusters with random ages and masses, selected to be representative of the objects in our real sample. 
This is just meant to provide some order of magnitude understanding of the impact different processes have on our result. 

The comparison of the predicted ionizing photon flux $\Qpred$ is shown as a corner plot in \cref{fig:corner_ionizing_photons}. 
Overall we find decent agreement between most models, but there are a few noticeable, systematic differences. 
The models with rotation predict higher ionizing photon fluxes, as expected from the study of \citet{Levesque+2012}. 
As for \textsc{bpass}, there are two distinct branches: during the first few \si{\mega\year} the agreement is good, but around \SI{6}{\mega\year} the difference between single stars and binaries becomes relevant. 

While the models can differ significantly from one another, those differences affect mostly a specific part of the samples, like older or low-mass clusters. 
For the objects that are young and massive, the differences are much smaller and all models return similar values. 
For the remainder of the paper, we use the \textsc{starburst99} model without rotation as our fiducial model.

\subsection{Matched catalogues}\label{sec:matched_catalogues}

Based on the one catalogue for the \HII regions and the two for the stars (clusters and associations), it is possible to define different matched catalogues. 
Here we describe the different samples that are used in this paper. 
\cref{fig:matched_catalogue_cutouts} illustrates examples for possible overlaps and \cref{tbl:catalogues} provides an overview of the different samples and their sizes.

\subsubsection{{\upshape\texttt{one-to-one}} sample}\label{sec:one-to-one}

The ideal scenario is a single \HII region that is powered by a single stellar population, as this enables us to study trends of the escape fraction with different nebula and stellar properties. 
This is the case for the catalogue that was published in \citetalias{Scheuermann+2023}. 
It consists of \HII regions and stellar associations that have a clear one-to-one match and contains \num{4169} objects in the full sample. 
This includes \SI{20.5}{\percent} of all \HII regions and \SI{25.6}{\percent} of the associations. 

For lower mass associations, the presence or absence of individual massive stars can have a significant impact on the measured fluxes \citep{Fouesneau+2010,Fouesneau+2012,Hannon+2019}. 
To alleviate stochastic sampling effects of the IMF, we also define a \texttt{robust one-to-one} subsample, consisting of objects that are more massive than \SI{e4}{\Msun}, with ages of \SI{8}{\mega\year} or less, and fully contained within the nebulae. 
This leaves us with \num{474} objects and constitutes our fiducial sample that we will use as a reference throughout the paper. 
They are spread across all galaxies apart from \galaxyname{NGC}{5068} (the lowest mass galaxy in our sample). 
Panel a) in \cref{fig:matched_catalogue_cutouts} shows an example for such an object.

\subsubsection{{\upshape\texttt{extended}} sample}

As laid out in \citetalias{Scheuermann+2023}, a considerable number of our detected \HII regions ($\SI{14.7}{\percent}$) overlap with more than one association. 
In this regard, we find \num{2981} \HII regions that contain \num{8871} stellar associations, but some of the associations overlap with another \HII region, making it difficult to map their ionizing photon flux to the correct nebula. 
We therefore limit ourselves to the \num{966} \HII regions where the associations do not overlap with another \HII region (a total of \num{2248} associations). 
Over three-quarters of them contain two associations and only a single one contains more than six associations. 
We refer to this sample as the \texttt{extended} sample and panel b) and c) in \cref{fig:matched_catalogue_cutouts} show postage stamp cutouts for this sample. 

Again we aim to define a robust subsample, however, in this case the definition is not so straightforward. 
We require that at least one association is young and massive ($>\SI{e4}{\Msun}$ and $\leq\SI{8}{\mega\year}$), and there are \num{707} of them in the initial catalogue. 
As for the overlap, we are less concerned about the old or light associations, but if a young and massive one only partially overlaps, we remove the object. 
This leaves us with \num{291} objects for our \texttt{robust extended} subsample. 
In comparison to the \texttt{one-to-one} sample, the \HII regions in the \texttt{extended} catalogue are brighter in $\HA$ by a factor of \num{5}.

\subsubsection{{\upshape\texttt{complexes}} sample}

As alluded to in \cref{sec:data:HII_regions}, difficulties in drawing the boundaries of the \HII regions can lead to large uncertainties for the measured fluxes. 
This is particularly problematic for the $\SI{34}{\percent}$ of our \HII regions that touch another ionized nebula. 

\begin{figure}
    \centering
    \includegraphics[width=\columnwidth]{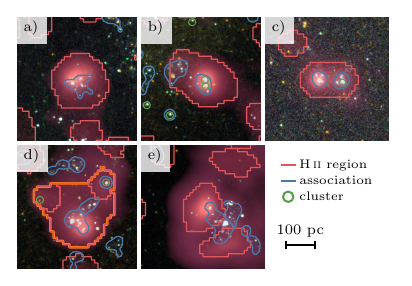}
    \caption[Overlap postage stamps.]{Examples for the overlap between \HII regions and their ionizing sources. 
    Each cutout shows three-colour composite images, based on the five available \textit{HST} bands, overlaid with the $\HA$ line emission of MUSE in red. 
    a) an \HII region with a fully contained association; b) an \HII region with a fully and a partially contained association and two compact clusters; c) an \HII region with two fully contained associations and a compact cluster; d) multiple \HII regions that form a single \HII region complex with multiple associations and clusters; e) an association that overlaps with two \HII regions.}
    \label{fig:matched_catalogue_cutouts}
\end{figure}

For this sample, we group bordering nebulae together and call them a complex. 
We find \num{9434} nebulae in \num{3019} complexes, but many of them contain ionized nebulae inconsistent with photoionization (e.g.\ SNRs or PNe), indicating that other processes may contribute to the $\HA$ emission. 
There are also complexes that do not contain an ionizing source while for some the ionizing source is shared with another complex (two complexes that do not touch, but share an association that overlaps with both as shown in panel e) of \cref{fig:matched_catalogue_cutouts}). 
Excluding all of them leaves \num{1124} complexes in our sample that contain \num{3722} \HII regions and \num{3923} associations. 
An example for such an object can be seen in panel d) of \cref{fig:matched_catalogue_cutouts}. 

For the robust subsample, we apply the same criteria as for the \texttt{robust extended} sample, i.e.\ at least one fully contained young and massive association and no partially overlapping young and massive ones. 
This leaves \num{317} objects in the \texttt{robust complexes} subsample.

\subsubsection{{\upshape\texttt{cluster}} sample}

In order to directly compare our results with existing studies that focused on star clusters instead of stellar associations \citep[e.g.][]{DellaBruna+2021,Teh+2023}, we also construct a sample focused on the overlap of \HII regions and star clusters. 
We use the Class 1, 2, and 3 clusters from \citet{Thilker+2022} and find \num{22632} clusters that overlap with \num{8905} \HII regions. 
Similar to the \texttt{extended} sample, having multiple clusters inside one \HII region is not an issue, and this constitutes our \texttt{cluster} sample. 
We also apply the same mass and age restrictions for the robust subsample (the overlap criterion does not apply as the clusters are treated as point sources), and find \num{2223} objects in the \texttt{robust cluster} subsample.

\begin{table}
    \centering
    \caption{The matched samples we define for this paper.}
    \begin{tabular}{
        l
        S[table-format=4]
        r
        p{3.5cm}
        }
        \toprule
        Name & \multicolumn{2}{c}{Sample size} & Description \\\midrule
        \texttt{one-to-one} & 4169 & (\num{474}) & \HII regions with a one-to-one match to a stellar association (from \citetalias{Scheuermann+2023}). \\ 
        \texttt{extended} & 966 & (\num{291}) & \HII regions that contain multiple stellar associations. \\ 
        \texttt{complexes} & 1124 & (\num{317}) & Multiple \HII regions that form a larger complex and contain one or more associations. \\ 
        \texttt{cluster} & 8905 & (\num{2223}) & \HII regions that contain one or more compact clusters. \\ 
        \bottomrule
    \end{tabular}
    \label{tbl:catalogues}
    \tablefoot{The size of the robust subsamples is in parenthesis.}
\end{table}

\section{Escape fractions}\label{sec:escape}

\begin{figure*}
    \centering
    \includegraphics[width=0.9\textwidth]{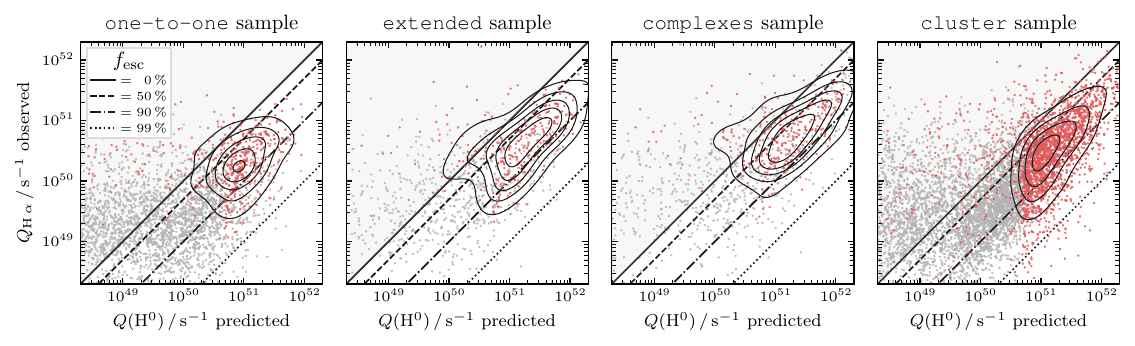}
    \caption[Comparison between the predicted ionizing photon flux $\Qpred$ and the observed value $\Qobs$ for individual objects.]{We compare the predicted ionizing photon flux $\Qpred$ to the observed values from the \HII region $\Qobs$. 
    The full samples are shown in grey and the robust subsamples in red. 
    The contours indicate the distribution of the robust sample and the diagonal lines correspond to different escape fractions.} 
    \label{fig:ionizing_photons_sample}
\end{figure*}

As is evident from the previous section, there are many assumptions involved in the calculation, but the choice of the stellar population model should only slightly alter the value of $\Qpred$ for the robust subsamples. 
Here we shift our focus to the different samples that were defined. 
By combining the values from the \HII regions to those from the stellar catalogues we estimate the percentage of leaking radiation.

\subsection{Ionizing photon fluxes for different samples}\label{sec:ionizing_photons}

We start by taking the samples defined in \cref{sec:matched_catalogues} and plot their observed ionizing photon flux from the \HII regions, $\Qobs$, against the one predicted from the stars, $\Qpred$. 
The result is shown in \cref{fig:ionizing_photons_sample} and the robust subsamples are marked in red. 

Looking on the \texttt{one-to-one sample}, we find that for more than half of the full sample, the ionizing photon production rate from the associations is not sufficient to explain the observed $\HA$ flux. 
However, most of those objects are either low-mass or only partially overlap. 
Focusing only on the previously defined \texttt{robust one-to-one} subsample, we find that $\SI{80}{\percent}$ of the \HII regions can be ionized by their associated stars. 
The stellar association catalogues created at different physical scales ($\SI{16}{\parsec}$ or $\SI{64}{\parsec}$, see also \citetalias{Scheuermann+2023}) provide similar results. 

For the other samples, we find similar trends, emphasising the importance of selecting robust subsamples. 
It also becomes clear that the different catalogues probe different regimes in terms of the intensity of the ionizing photon flux. 
Unsurprisingly, the \texttt{extended sample} and the \texttt{complexes sample} have ionizing photon fluxes that are on average more than four times as large as the \texttt{one-to-one} sample.

\subsection{Distribution of the escape fraction}\label{sec:escape_fractions}

\begin{figure}
    \centering
    \includegraphics[width=0.75\columnwidth]{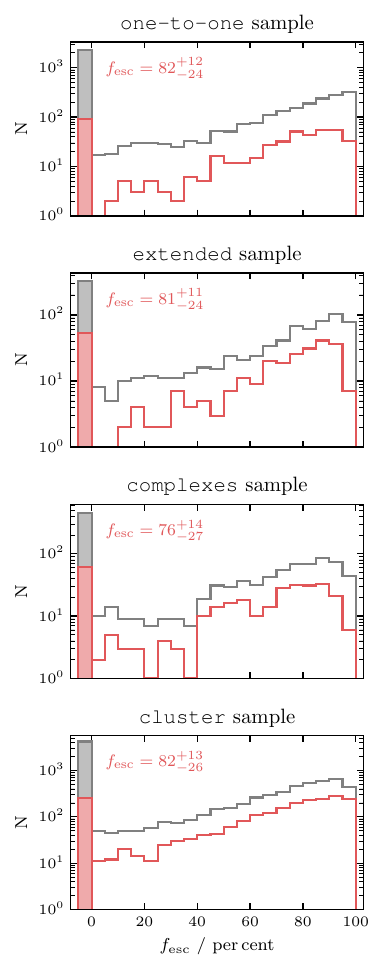}
    \caption[Histogram of the observed escape fractions for the different samples.]{Histograms of the observed escape fractions for the different samples. 
    All regions with negative escape fractions are grouped together in the shaded left bar. 
    The full sample is shown in gray and the robust subsample and its median in red.} 
    \label{fig:escape_fractions}
\end{figure}

With $\Qpred$ from the stars and $\Qobs$ from the \HII regions, we use \cref{eq:escape_fraction} to compute the percentage of escaping ionizing radiation. 
As already obvious from \cref{fig:ionizing_photons_sample}, for a substantial number of objects, the ionizing photon flux from the stars is not sufficient to ionize their matched \HII region. 
As a result, some objects have a negative escape fraction, which makes it somewhat ambiguous how to determine a representative escape fraction for the sample as a whole. 
Excluding all regions with negative $\fesc$ leads to a higher average, while including values that are clearly unphysical also does not make sense. 
Our approach is a compromise to account for objects that have a positive escape fraction within their uncertainties. 
We assume that the escape fraction for each object follows a normal distribution and sum up the individual distributions
\begin{equation}\label{eq:fesc_dist}
    p(x) = \sum_{i=1}^n \frac{1}{\sqrt{2\pi}\, \sigma_{f_{\mathrm{esc}},i}} \exp \left( \frac{(\mu_{f_{\mathrm{esc}},i} - x)^2 }{2\, \sigma_{f_{\mathrm{esc}},i}^2} \right) . 
\end{equation}
We then compute the median (50th percentile) and the $1\sigma$ interval (16th and 84th percentile) in the range of valid values (\SIrange{0}{100}{\percent}). 
The results are shown as histograms in \cref{fig:escape_fractions} and the majority has an escape fraction larger than $\SI{50}{\percent}$. 
In the case of the robust subsamples, independent of the source of ionization, we measure similar medians, ranging from \SIrange{76}{82}{\percent}. 
The medians of the full samples are always within a few \si{\percent} of those values. 
The number of objects with negative escape fractions varies between \SIrange[range-phrase = { and }]{12}{19}{\percent} and they are grouped together in the left bar.

\subsection{What about negative escape fractions?}\label{sec:neg_fesc}

There is a considerable number of objects with a negative escape fraction, so here we take a closer look at what may cause this. 
The objects with a negative escape fraction in the \texttt{robust one-to-one} subsample are consistent with $\fesc=0$ within their uncertainties. 
Looking at the contributors to the absolute uncertainty, the dominant factor is the age, followed slightly behind by the mass while the error on the $\HA$ flux is mostly negligible. 
It is therefore not surprising that most objects with a negative escape fraction are \SI{4}{\mega\year} or older and have a median uncertainty of \SI{2}{\mega\year} which makes them particularly susceptible that their predicted flux is wrong. 
An overestimation of the age could account for an underestimation of $\Qpred$ and hence resolve most of the negative escape fractions. 

Another possibility is that we are missing some ionizing sources all together. 
To explore this, we visually inspect the NUV images of the 92 objects with negative escape fraction and look for missed peaks inside the \HII regions. 
Only about $\SI{10}{\percent}$ of them show peaks that were not included by the source detection of the associations. 

\begin{figure}[htb]
    \centering
    \includegraphics[width=0.9\columnwidth]{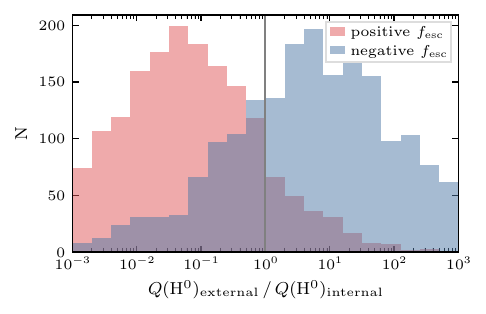}
    \caption{Contribution of external ionizing sources. 
    For \HII regions with negative escape fractions, the average contribution from nearby associations, not overlapping with the nebula, is far greater than for those with positive escape fractions.} 
    \label{fig:external_ionizing_sources}
\end{figure}

Along these lines, the cloud might also be ionized by stars that fall outside of the \HII region. 
To follow up on this point, for each \HII region, we compile a list of associations that are within $\SI{10}{\arcsec}$ (this corresponds to \SIrange{250}{950}{\parsec}, measured from boundary to boundary). 
We then take the flux of each of those nearby associations (excluding associations that are fully contained in another \HII region) and scale the estimated ionizing photons that could be absorbed by the \HII region, assuming the flux is emitted isotropically. 
For each \HII region, we compare the ionizing photon flux that originates from external sources to the one coming from overlapping sources. 
The result is shown in \cref{fig:external_ionizing_sources}. 
We find that in the subsample with negative escape fraction the estimated contribution from external ionizing photons is about a factor of \num{10} more than the internal contribution, which is more than a factor \num{100} larger than what is found in the subsample with positive escape fraction. 
However, even after accounting for these external contributions, the majority of the objects with negative escape fraction remain negative. 
Nevertheless, it shows that the escape fraction is underestimated if external sources are not taken into account. 

So far we have mainly focused on the stellar side, but the $\Qobs$ estimated from the \HII regions is also subject to uncertainties. 
Given that the $\HA$ fluxes are directly measured and have a high S/N, the stated errors are significantly smaller for $\Qobs$ than for $\Qpred$, but this does not include uncertainties due to choices like where we define the boundaries of the regions. 
Even a small change in the size can significantly alter the measured flux. 
This is especially true for closely clustered and unresolved \HII regions, where we might attribute the measured flux to a neighbouring object. 
Increasing or decreasing the boundary by just one pixel radially (this corresponds to \SIrange{5}{19}{\parsec}, depending on the distance to the galaxy) will alter the measured flux on average by $\pm\SI{40}{\percent}$ for the full \HII regions sample, by $\pm\SI{30}{\percent}$ for the matched objects, and by $\pm\SI{18}{\percent}$ for the robust subsample. 
However, the flux measured from the \HII regions is only roughly a fifth of the one predicted from the stars, so although the uncertainty itself is high, it typically changes the escape fraction by less than \SI{5}{\percent}. 

In addition, in particular for faint \HII regions, the DIG might contribute a majority of $\HA$ emission, though our brighter robust sample should only be impacted at a $\sim\SI{20}{\percent}$ level \citep{Belfiore+2022,Congiu+2023}.

\subsection{Photoionization budget of entire galaxies}\label{sec:ionization_budget_galaxies}

\begin{table*}
    \centering
    \caption[Ionizing photon budget for entire galaxies.]{Ionizing photon budget for entire galaxies.}
    \begin{tabular}{
        l
        S[table-format=4.2]@{\,\( \pm \)\,}
        S[table-format=1.6]
        S[table-format=4.2]@{\,\( \pm \)\,}
        S[table-format=1.6]
        S[table-format=3.1]@{\,\( \pm \)\,}
        S[table-format=2.2]
        S[table-format=4.2]@{\,\( \pm \)\,}
        S[table-format=1.6]
        S[table-format=3.2]@{\,\( \pm \)\,}
        S[table-format=4.2]
        }
        \toprule
        & \multicolumn{2}{c}{\HII regions} & \multicolumn{4}{c}{Stellar associations} & \multicolumn{4}{c}{Star cluster} \\
        \cmidrule(lr){2-3}\cmidrule(lr){4-7}\cmidrule(lr){8-11}
        Name & \multicolumn{2}{r}{$\log \left( \Qobs \,/\,\si{\per\second}\right)$} & \multicolumn{2}{c}{$\log \left( Q(\mathrm{H}^0)\,/\,\si{\per\second}\right)$} & \multicolumn{2}{c}{$f_\mathrm{esc}$ / per\,cent} & \multicolumn{2}{c}{$\log \left( Q(\mathrm{H}^0)\,/\,\si{\per\second}\right)$} & \multicolumn{2}{c}{$f_\mathrm{esc}$ / per\,cent} \\\midrule
        \galaxyname{IC}{5332} & 51.56 & 0.06 & 52.38 & 0.04 & 84.9 & 2.5 & 52.13 & 0.36 & 73.0 & 22.5 \\
        \galaxyname{NGC}{0628} & 52.50 & 0.04 & 53.29 & 0.01 & 83.5 & 1.6 & 53.05 & 0.58 & 71.7 & 37.7 \\
        \galaxyname{NGC}{1087} & 53.11 & 0.06 & 53.65 & 0.04 & 71.4 & 4.6 & 53.55 & 0.15 & 63.7 & 13.0 \\
        \galaxyname{NGC}{1300} & 52.70 & 0.04 & 52.97 & 0.06 & 45.5 & 9.2 & 53.25 & 1.98 & 71.5 & 130.1 \\
        \galaxyname{NGC}{1365} & 53.15 & 0.14 & 53.43 & 0.06 & 47.5 & 18.4 & 53.48 & 0.57 & 52.8 & 64.0 \\
        \galaxyname{NGC}{1385} & 53.34 & 0.05 & 53.70 & 0.04 & 56.3 & 6.4 & 53.61 & 0.04 & 46.8 & 7.3 \\
        \galaxyname{NGC}{1433} & 52.26 & 0.04 & 52.71 & 0.02 & 64.0 & 4.0 & 53.00 & 0.14 & 81.4 & 7.1 \\
        \galaxyname{NGC}{1512} & 52.29 & 0.05 & 52.62 & 0.03 & 53.2 & 6.3 & 53.59 & 0.17 & 95.0 & 5.6 \\
        \galaxyname{NGC}{1566} & 53.44 & 0.06 & 53.99 & 0.02 & 71.7 & 4.0 & 53.85 & 0.20 & 61.4 & 17.8 \\
        \galaxyname{NGC}{1672} & 53.26 & 0.05 & 53.72 & 0.02 & 65.5 & 4.1 & 53.67 & 0.42 & 61.1 & 37.4 \\
        \galaxyname{NGC}{2835} & 52.56 & 0.05 & 53.14 & 0.03 & 74.1 & 3.6 & 53.18 & 0.44 & 76.1 & 24.2 \\
        \galaxyname{NGC}{3351} & 51.99 & 0.04 & 52.95 & 0.05 & 89.0 & 1.6 & 52.72 & 0.19 & 81.5 & 8.3 \\
        \galaxyname{NGC}{3627} & 53.42 & 0.04 & 53.75 & 0.05 & 53.9 & 7.0 & 53.81 & 0.29 & 59.7 & 27.5 \\
        \galaxyname{NGC}{4254} & 53.47 & 0.02 & 54.06 & 0.03 & 74.1 & 2.2 & 54.10 & 0.04 & 76.6 & 2.6 \\
        \galaxyname{NGC}{4303} & 53.60 & 0.03 & 54.13 & 0.02 & 70.0 & 2.6 & 54.17 & 0.07 & 72.9 & 4.9 \\
        \galaxyname{NGC}{4321} & 53.05 & 0.04 & 53.77 & 0.02 & 80.8 & 2.1 & 53.69 & 0.05 & 76.9 & 3.3 \\
        \galaxyname{NGC}{4535} & 52.73 & 0.04 & 53.27 & 0.05 & 71.0 & 4.2 & 53.34 & 0.16 & 75.4 & 9.6 \\
        \galaxyname{NGC}{5068} & 52.20 & 0.05 & 52.63 & 0.04 & 62.9 & 5.6 & 52.67 & 0.12 & 65.7 & 10.8 \\
        \galaxyname{NGC}{7496} & 52.62 & 0.06 & 53.23 & 0.03 & 75.8 & 3.9 & 53.20 & 0.05 & 74.0 & 4.7 \\\midrule
        Total & 54.36 & 0.02 & 54.87 & 0.01 & 68.9 & 1.3 & 54.88 & 0.07 & 69.6 & 5.0 \\
        \bottomrule
    \end{tabular}
    \tablefoot{We use the \HII regions from \citet{Santoro+2022} and \citet{Groves+2023} and either the stellar associations ($\SI{32}{\parsec}$, NUV) from \citet{Larson+2023}, or the compact star clusters from \citet{Thilker+2022}.}
    \label{tbl:ionizing_photons_galaxies}
\end{table*}

As we previously saw, it is difficult to correctly assign each \HII region to their source of ionizing radiation. 
A careful selection is required in order to include all real sources, while excluding any false ones. 
Another approach would be to simply consider all objects together in order to avoid problems with the matching. 
And as we saw in \cref{fig:external_ionizing_sources}, sources outside the \HII region can have a significant contribution to the ionizing photon budget as well. 
We therefore perform this analysis on all \HII regions and associations in the overlapping coverage of MUSE and \textit{HST}. 
Since we have no way of excluding \HII regions that are ionized by older or low-mass associations, this analysis also has to include all associations and not just the \texttt{robust} subsample. 
One potentially problematic area are the galactic centres, as some galaxies possess active nuclei. 
We therefore use the environment masks from \citet{Querejeta+2021} to mask out the centres of the galaxies (this eliminates \num{286} \HII regions and \num{690} associations from our sample). 

\begin{figure}
    \centering
    \includegraphics[width=0.8\columnwidth]{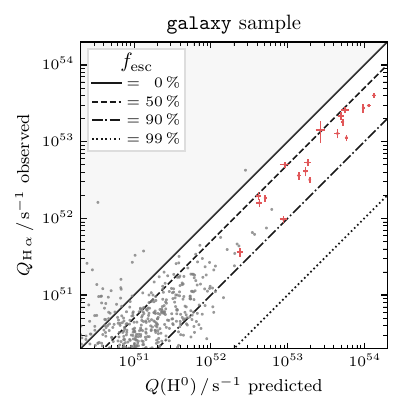}
    \caption[Comparison between the predicted ionizing photon flux $\Qpred$ and the observed value $\Qobs$ for entire galaxies.]{Same as \cref{fig:ionizing_photons_sample} but with the ionizing photon budget for entire galaxies. 
    The values for the galaxies are listed in \cref{tbl:ionizing_photons_galaxies} and include all \HII regions and stellar associations in the overlapping coverage. 
    For reference, the values of individual objects from the \texttt{robust one-to-one} and \texttt{robust extended} sample are plotted as grey dots.}
    \label{fig:ionizing_photons_by_galaxy}
\end{figure}

We sum up the observed ionizing photon flux across all \HII regions that fall inside the \textit{HST} coverage, and measure $\Qobs=\SI{2.3e+54}{\per\second}$ across all galaxies. 
We then sum up the predicted ionizing photon flux from all associations inside the MUSE coverage, and find $\Qpred=\SI{7.2e+54}{\per\second}$, yielding an escape fraction of $\fesc=\SI{68.9\pm1.3}{\percent}$. 
Similarly, we can sum up the flux from all star clusters, resulting in a value of $\Qpred=\SI{7.4e54}{\per\second}$ and an escape fraction of $\fesc=\SI{69.6\pm5.0}{\percent}$. 
The values for the individual galaxies are listed in \cref{tbl:ionizing_photons_galaxies} and displayed in \cref{fig:ionizing_photons_by_galaxy}. 

From this approach, the upper end of the luminosity function should be fully sampled for both the associations and \HII regions, but there are differences in the sensitivity at the lower end. 
When looking at the cumulative luminosity function (see also \cref{sec:appendix_completeness} for more details), we see that it is dominated by the brightest objects. 
\HII regions fainter than $\SI{e38}{\erg\per\second}$ contain less than $\SI{10}{\percent}$ of the total ionizing photon flux (this is already ten times brighter than the completeness limit estimated by \citealt{Santoro+2022}). 
This means that completeness should not be an issue for the \HII regions. 
As for the stars, they are even more dominated by the upper end of the luminosity function. 

Other minor adjustments to our approach have only a negligible impact on our global result. 
If we have excluded some real \HII regions with the BPT cuts, thereby artificially reducing the observed ionizing photon flux and increasing the escape fraction, we estimate that these only contribute around $\sim\SI{10}{\percent}$ to the total $\HA$ flux in all nebulae. 
This would only decrease the resulting escape fraction by a few \si{\percent}. 
There are also some compact clusters that are not contained inside an association. 
Including them in our analysis raises the predicted ionizing photon flux only slightly and the escape fraction only increases by $\sim\!\SI{3}{\percent}$. 

Overall, both our global approach and our resolved approach return comparatively high values of $\SI{70}{\percent}$ and above, providing confidence in our overall result for this large statistical sample of stellar sources and \HII regions. 

\subsection{Trends with nebulae properties}

\begin{figure*}
    \centering
    \includegraphics[width=\textwidth]{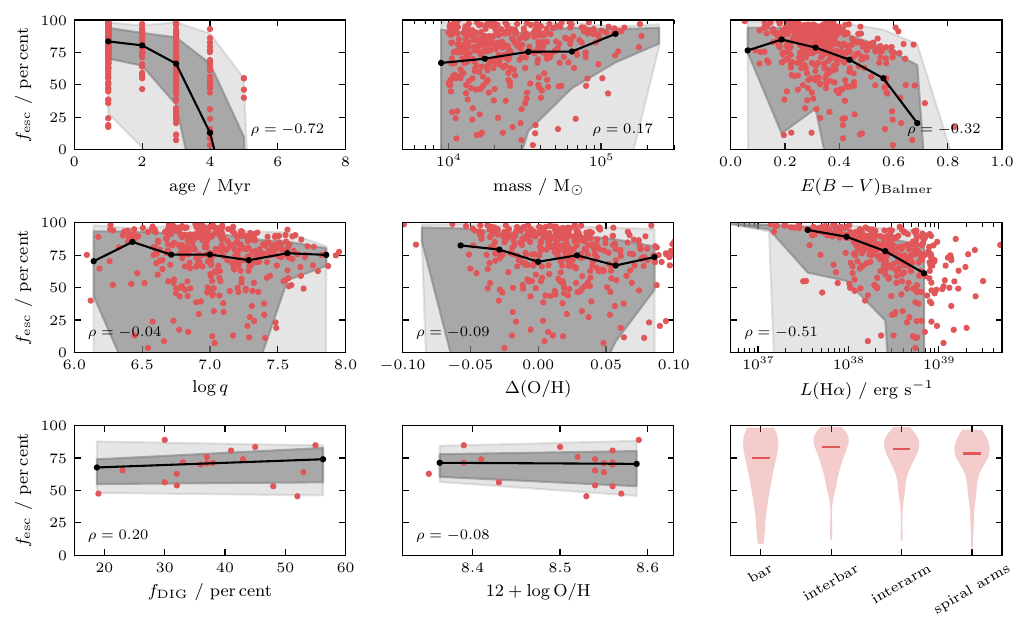}
    \caption[Trends between different nebula properties and the escape fraction.]{Trends between different nebula properties and the escape fraction. 
    Shown are the age and mass of the stellar associations, the reddening $E(B-V)_\mathrm{Balmer}$, the ionization parameter $\log q$, the local metallicity offset $\DeltaOH$ of the \HII region, and the luminosity $L(\HA)$. 
    The last row shows the global escape fraction (see \cref{sec:ionization_budget_galaxies}) versus the fraction of DIG emission, the global escape fraction versus the average metallicity of the galaxy, and the escape fraction for different environments. 
    The black lines represent the median values in each bin and the gray shaded areas indicate the 68 and 98 percentiles.} 
    \label{fig:trends_with_fesc}
\end{figure*}

The escape fraction is expected to change as the cloud evolves \citep{Dayal+2018} and as a function of local conditions. 
For example, one might expect higher $\fesc$ in lower density environments or older \HII regions, as the gas is already partially cleared. 
In \cref{fig:trends_with_fesc} we present trends with selected properties. 
For this analysis we use the \texttt{robust one-to-one} subsample. 
In general, the large scatter in $\fesc$ makes it challenging to identify strong correlations. 

Contrary to our expectations, we find that \HII regions containing younger associations are leakier \citep[similar to the results reported by][]{Teh+2023}. 
As a function of the stellar mass of the associations, the scatter is too large to identify any trends. 
In terms of the dust reddening, as traced by the Balmer decrement, we observe an anti-correlation, consistent with the picture that a deeply embedded cloud leaks less radiation than a more permeable one. 
However, using the stellar $E(B-V)$ we do not find a strong correlation. 
For the ionization parameter, $\log q$, and the local metallicity offset, $\Delta (\mathrm{O}/\mathrm{H})$, we do not find any correlation. 
Lastly, for the individual \HII regions, we look at the $\HA$ luminosity and find an anti-correlation with the escape fraction, similar to \citet{Teh+2023}. 

Based on our global measurements of $\fesc$, we also search for galaxy-wide trends. 
The escaping photons are thought to ionize the DIG and hence we expect a positive correlation. 
However, we see no trend between our observed $\fesc$ and the fraction of the DIG $f_\mathrm{DIG}$ measured by \citet{Belfiore+2022}, whereby our values are also generally larger. 
Another property is the global metallicity $12+\log \mathrm{O}/\mathrm{H}$ \citep[taken from][]{Kreckel+2020}, but here too we see no dependency. 

Finally, we look at the variations between different galactic environments within each galaxy, considering separately the bar, interbar, interarm, and spiral arm environments. 
While overall they are very similar, with medians ranging from \SIrange{75}{83}{\percent}, the \HII regions in the bar have slightly lower escape fractions.

\subsection{Comparison with existing studies}

The large sample of \HII regions and stellar sources that we use to measure the escape fraction results in a wide span of values. 
Looking at their distribution, it becomes evident that it is deceptive to quote a single value. 
The values published in the literature reinforce this as they also cover a wide range, depending on the conditions of the studied objects and the techniques used for the analysis. 

We therefore focus our comparison on similar studies that also match observed \HII regions to their stellar counterparts in nearby galaxies. 
Most of our measurements of the escape fraction exceeds \SI{60}{\percent}, which itself falls at the upper end of values present in the literature (\SIrange{9}{67}{\percent}) of comparable studies like \citet{McLeod+2020} or \citet{DellaBruna+2021}. 
Even for independent calculations done in exactly the same galaxy (\galaxyname{NGC}{0628} by \citealt{Teh+2023}; see Appendix \ref{sec:appendix_slug}), our approach returns much larger values for $\fesc$. 
While it is clear that specific modelling and measurement approaches can introduce large scatter, it is not obvious that there is any clear systematic difference from our approach compared to what has been previously done. 
We aim to highlight here that these modelling uncertainties remain dominant while attempting to quantitively constrain the escape fraction and below we emphasize a number of issues that stand out. 

The large size of our sample prompted us to introduce certain cuts to define a \texttt{robust} subsample where we remove objects we deem unreliable. 
The stellar component must be fully contained within the \HII region, its mass must be larger than \SI{e4}{\Msun} to ensure that the IMF is fully sampled, and its age must be $\leq\SI{8}{\mega\year}$ because the ionized cloud should not exist much longer than a few \si{\mega\year} \citep{Hollyhead+2015,Kim+2021,Hannon+2022}. 
This procedure is certainly more objective than picking the objects by hand, but it might still be biased towards certain physical parameters. 
Partly due to this selection, the regions in our sample are rather young with a median age of \SI{2}{\mega\year}. 
Nevertheless, the age distribution seems physically reasonable, as according to \citet{Kim+2023}, \SI{50}{\percent} of the LyC photons are already emitted by this age. 
Around this age, the ionizing photon flux is strongly dependent on the age and hence even a small uncertainty of only \SI{1}{\mega\year} can cut the value for $\Qpred$ in half. 
This is also reflected in the propagated error of $\fesc$, for which this is the biggest contributor. 

Then there are also other issues whose uncertainties are difficult to incorporate. 
As mentioned in \cref{sec:data:HII_regions}, a small change to the size of a region can alter the flux considerably and this is not accounted for in the stated error. 
When manually inspecting some of the outliers, it becomes clear that especially the larger \HII regions are often complicated complexes that are difficult to segment into physically meaningful pieces. 
Furthermore, we do not consider external sources, which can have a major impact on small \HII regions as shown in \cref{sec:neg_fesc}. 

One aspect that we have not yet addressed is the impact of dust. 
This is a complex issue, and while some simulations are able to account for the ionizing photons that are directly absorbed by the dust, we are only able to correct the $\HA$ fluxes for reddening. 
According to \citet{Kim+2019} a double-digit percentage might be lost this way. 
This cannot explain the difference to other observations, which have generally followed the same approach as we have, but it could help explain the higher values that we observe when compared to simulations.

\section{Conclusion}\label{sec:conclusion_escape}

We combine catalogues of \HII regions with catalogues of young stellar associations and star clusters to measure the escape fraction across an unprecedentedly large sample of thousands of regions across 19 nearby galaxies. 
This dataset enables a statistical view of how much ionizing radiation escapes from the local \HII region environment out into the galaxy disk. 

To perform the calculation of $\fesc$, we compare $\Qobs$, derived from the extinction-corrected $\HA$ luminosity, and $\Qpred$, derived from the associated stellar counterpart. 
After comparing different methods of matching stars and gas, we focus our analysis on regions with a one-to-one match between a single \HII region and a single stellar association. 
To mitigate the impact of stochastic sampling, we establish a \texttt{robust} subsample of \num{474} massive ($>\SI{e4}{\Msun}$) and young ($\leq\SI{8}{\mega\year}$) stellar associations that are fully contained in the nebula. 

\begin{enumerate}[label=(\roman*)]
    \item From the \texttt{robust one-to-one} subsample, we measure a comparatively high value of $\fesc=\uncertainty{82}{12}{24}\;\si{\percent}$. \\[-0.3cm]
    \item We derive similar escape fractions with different crossmatching methods or when using compact clusters instead of the loosely bound associations. 
    \item The escape fraction depends heavily on the derived stellar parameters and the boundaries of the \HII regions. 
   \item For many \HII regions, a significant fraction of the ionizing photon budget comes from stars that reside outside their boundaries. 
    \item If we take all stellar associations and \HII regions together, to account for the entire ionizing photon budget in the galaxies, we measure slightly lower but consistent values. 
    \item We do not identify any pronounced trends between $\fesc$ and local or global galaxy properties. 
\end{enumerate}

This work demonstrates that a large fraction of ionizing photons can escape their immediate \HII region bubble, and be available to power the surrounding pervasive DIG observed in local galaxies \citep{Haffner+2009, Belfiore+2022}. 

Overall, more work is needed on precise determination of the stellar ionizing photon output before it is possible to establish more concrete connections between changes in $\fesc$ and local conditions in the ISM. 

We expect that significant progress on parametrising the properties of the stellar clusters will come from the inclusion of additional infrared \textit{JWST} bands, and the use of priors based on the presence of $\HA$ emission \citep{Whitmore+2025}. 
In addition, the recent availability of new \textit{HST} $\HA$ high resolution maps \citep{Chandar+2025} is enabling more detailed crossmatching between ionized gas and stellar populations at matched resolution for the brightest, most compact regions \citep{Barnes+2026}. 
These maps already reveal the ways in which the clustering and hierarchical structuring of young stars make simple one-to-one matching of ionized gas and stars extremely challenging.

\begin{acknowledgements}

This work has been carried out as part of the PHANGS collaboration. 
Based on observations from the PHANGS-MUSE program, collected at the European Southern Observatory under ESO programmes 094.C-0623 (PI: Kreckel), 095.C-0473, 098.C-0484 (PI: Blanc), 1100.B-0651 (PHANGS-MUSE; PI: Schinnerer), as well as 094.B-0321 (MAGNUM; PI: Marconi), 099.B-0242, 0100.B-0116, 098.B-0551 (MAD; PI: Carollo) and 097.B-0640 (TIMER; PI: Gadotti). 
In addition, this research is based on observations made with the NASA/ESA Hubble Space Telescope obtained from the Space Telescope Science Institute, which is operated by the Association of Universities for Research in Astronomy, Inc., under NASA contract NAS 5–26555. 
Support for Program number 15654 was provided through a grant from the STScI under NASA contract NAS5-26555. 

FS and KK gratefully acknowledge funding from the Deutsche Forschungsgemeinschaft (DFG, German Research Foundation) in the form of an Emmy Noether Research Group (grant number KR4598/2-1, PI Kreckel) and the European Research Council’s starting grant ERC StG-101077573 ``ISM-METALS''). 
MB acknowledges support from the ANID BASAL project FB210003. 
This work was supported by the French government through the France 2030 investment plan managed by the National Research Agency (ANR), as part of the Initiative of Excellence of Université Côte d’Azur under reference number ANR-15-IDEX-01. 
OE acknowledges funding from the Deutsche Forschungsgemeinschaft (DFG, German Research Foundation) -- project-ID 541068876. 
SCOG and RSK acknowledge financial support from the European Research Council via the ERC Synergy Grant ``ECOGAL'' (project ID 855130) and from the German Excellence Strategy via the Heidelberg Cluster of Excellence (EXC 2181 - 390900948) ``STRUCTURES''. 
JEMD acknowledges support from project UNAM DGAPA-PAPIIT IG 101025, Mexico. 
MC and LR gratefully acknowledge funding from the DFG through an Emmy Noether Research Group (grant number CH2137/1-1). 

This research made use of \textsc{astropy} \citep{Astropy+2013,Astropy+2018,Astropy+2022}, \textsc{numpy} \citep{Harris+2020}, and \textsc{matplotlib} \citep{Hunter+2007}. 
\end{acknowledgements}

\bibliographystyle{aa}
\bibliography{paper}

@Article{Kewley+2006,
  author  = {{Kewley}, Lisa J. and {Groves}, Brent and {Kauffmann}, Guinevere and {Heckman}, Tim},
  journal = {\mnras},
  title   = {{The host galaxies and classification of active galactic nuclei}},
  year    = {2006},
  month   = nov,
  number  = {3},
  pages   = {961-976},
  volume  = {372},
  adsurl  = {https://ui.adsabs.harvard.edu/abs/2006MNRAS.372..961K},
  doi     = {10.1111/j.1365-2966.2006.10859.x},
  eprint  = {astro-ph/0605681},
}

@Article{McLeod+2020,
  author  = {{McLeod}, Anna F. and {Kruijssen}, J.~M. Diederik and {Weisz}, Daniel R. and {Zeidler}, Peter and {Schruba}, Andreas and {Dalcanton}, Julianne J. and {Longmore}, Steven N. and {Chevance}, M{\'e}lanie and {Faesi}, Christopher M. and {Byler}, Nell},
  journal = {\apj},
  title   = {{Stellar Feedback and Resolved Stellar IFU Spectroscopy in the Nearby Spiral Galaxy NGC 300}},
  year    = {2020},
  month   = mar,
  number  = {1},
  pages   = {25},
  volume  = {891},
  adsurl  = {https://ui.adsabs.harvard.edu/abs/2020ApJ...891...25M},
  doi     = {10.3847/1538-4357/ab6d63},
  eid     = {25},
  eprint  = {1910.11270},
}

@Article{Kauffmann+2003,
  author  = {{Kauffmann}, G. and {Heckman}, T.~M. and {Tremonti}, C. and {Brinchmann}, J. and {Charlot}, S. and {White}, S.~D.~M. and {Ridgway}, S.~E. and {Brinkmann}, J. and {Fukugita}, M. and {Hall}, P.~B. and {Ivezi{\'c}}, {\v Z}. and {Richards}, G.~T. and {Schneider}, D.~P.},
  journal = {\mnras},
  title   = {{The host galaxies of active galactic nuclei}},
  year    = {2003},
  month   = dec,
  pages   = {1055-1077},
  volume  = {346},
  adsurl  = {https://ui.adsabs.harvard.edu/abs/2003MNRAS.346.1055K},
  doi     = {10.1111/j.1365-2966.2003.07154.x},
  eprint  = {astro-ph/0304239},
}

@Article{Barnes+2026,
  author  = {{Barnes}, A.~T. and {Chandar}, R. and {Kreckel}, K. and {Belfiore}, F. and {Pathak}, D. and {Thilker}, D. and {Leroy}, A.~K. and {Groves}, B. and {Glover}, S.~C.~O. and {McClain}, R. and {Amiri}, A. and {Bazzi}, Z. and {Boquien}, M. and {Congiu}, E. and {Dale}, D.~A. and {Egorov}, O.~V. and {Emsellem}, E. and {Grasha}, K. and {Gonzalez Lobos}, J. and {Henny}, K. and {He}, H. and {Indebetouw}, R. and {Lee}, J.~C. and {Li}, J. and {Liang}, F.-H. and {Larson}, K. and {Maschmann}, D. and {Meidt}, S.~E. and {Eduardo M{\'e}ndez-Delgado}, J. and {Neumann}, J. and {Pan}, H.-A. and {Querejeta}, M. and {Rosolowsky}, E. and {Sarbadhicary}, S.~K. and {Scheuermann}, F. and {{\'U}beda}, L. and {Williams}, T.~G. and {Weinbeck}, T.~D. and {Whitmore}, B. and {Wofford}, A. and {PHANGS Collaborationn}},
  journal = {\aap},
  title   = {{The PHANGS-MUSE/HST-H{\ensuremath{\alpha}} nebulae catalogue: Parsec-scale resolved structure, physical conditions, and stellar associations across nearby galaxies}},
  year    = {2026},
  month   = feb,
  pages   = {A95},
  volume  = {706},
  adsurl  = {https://ui.adsabs.harvard.edu/abs/2026A&A...706A..95B},
  doi     = {10.1051/0004-6361/202555751},
  eid     = {A95},
  eprint  = {2510.11778},
}

@Article{Martin+2005,
  author  = {{Martin}, D. Christopher and {Fanson}, James and {Schiminovich}, David and {Morrissey}, Patrick and {Friedman}, Peter G. and {Barlow}, Tom A. and {Conrow}, Tim and {Grange}, Robert and {Jelinsky}, Patrick N. and {Milliard}, Bruno and {Siegmund}, Oswald H.~W. and {Bianchi}, Luciana and {Byun}, Yong-Ik and {Donas}, Jose and {Forster}, Karl and {Heckman}, Timothy M. and {Lee}, Young-Wook and {Madore}, Barry F. and {Malina}, Roger F. and {Neff}, Susan G. and {Rich}, R. Michael and {Small}, Todd and {Surber}, Frank and {Szalay}, Alex S. and {Welsh}, Barry and {Wyder}, Ted K.},
  journal = {\apjl},
  title   = {{The Galaxy Evolution Explorer: A Space Ultraviolet Survey Mission}},
  year    = {2005},
  month   = jan,
  number  = {1},
  pages   = {L1-L6},
  volume  = {619},
  adsurl  = {https://ui.adsabs.harvard.edu/abs/2005ApJ...619L...1M},
  doi     = {10.1086/426387},
  eprint  = {astro-ph/0411302},
}

@Article{Zurita+2000,
  author  = {{Zurita}, A. and {Rozas}, M. and {Beckman}, J.~E.},
  journal = {\aap},
  title   = {{The origin of the ionization of the diffuse interstellar medium in spiral galaxies. I. Photometric measurements}},
  year    = {2000},
  month   = nov,
  pages   = {9-28},
  volume  = {363},
  adsurl  = {https://ui.adsabs.harvard.edu/abs/2000A&A...363....9Z},
}

@Article{Niederhofer+2016,
  author  = {{Niederhofer}, F. and {Hilker}, M. and {Bastian}, N. and {Ercolano}, B.},
  journal = {\aap},
  title   = {{Constraining the escape fraction of ionizing photons from H II regions within NGC 300: A concept paper}},
  year    = {2016},
  month   = jul,
  pages   = {A47},
  volume  = {592},
  adsurl  = {https://ui.adsabs.harvard.edu/abs/2016A&A...592A..47N},
  doi     = {10.1051/0004-6361/201527819},
  eid     = {A47},
  eprint  = {1606.01144},
}

@Article{Hannon+2019,
  author  = {{Hannon}, Stephen and {Lee}, Janice C. and {Whitmore}, B.~C. and {Chand ar}, R. and {Adamo}, A. and {Mobasher}, B. and {Aloisi}, A. and {Calzetti}, D. and {Cignoni}, M. and {Cook}, D.~O. and {Dale}, D. and {Deger}, S. and {Della Bruna}, L. and {Elmegreen}, D.~M. and {Gouliermis}, D.~A. and {Grasha}, K. and {Grebel}, E.~K. and {Herrero}, A. and {Hunter}, D.~A. and {Johnson}, K.~E. and {Kennicutt}, R. and {Kim}, H. and {Sacchi}, E. and {Smith}, L. and {Thilker}, D. and {Turner}, J. and {Walterbos}, R.~A.~M. and {Wofford}, A.},
  journal = {\mnras},
  title   = {{H {\ensuremath{\alpha}} morphologies of star clusters: a LEGUS study of H II region evolution time-scales and stochasticity in low-mass clusters}},
  year    = {2019},
  month   = dec,
  number  = {4},
  pages   = {4648-4665},
  volume  = {490},
  adsurl  = {https://ui.adsabs.harvard.edu/abs/2019MNRAS.490.4648H},
  doi     = {10.1093/mnras/stz2820},
  eprint  = {1910.02983},
}

@Article{Chisholm+2022,
  author  = {{Chisholm}, J. and {Saldana-Lopez}, A. and {Flury}, S. and {Schaerer}, D. and {Jaskot}, A. and {Amor{\'\i}n}, R. and {Atek}, H. and {Finkelstein}, S.~L. and {Fleming}, B. and {Ferguson}, H. and {Fern{\'a}ndez}, V. and {Giavalisco}, M. and {Hayes}, M. and {Heckman}, T. and {Henry}, A. and {Ji}, Z. and {Marques-Chaves}, R. and {Mauerhofer}, V. and {McCandliss}, S. and {Oey}, M.~S. and {{\"O}stlin}, G. and {Rutkowski}, M. and {Scarlata}, C. and {Thuan}, T. and {Trebitsch}, M. and {Wang}, B. and {Worseck}, G. and {Xu}, X.},
  journal = {\mnras},
  title   = {{The far-ultraviolet continuum slope as a Lyman Continuum escape estimator at high redshift}},
  year    = {2022},
  month   = dec,
  number  = {4},
  pages   = {5104-5120},
  volume  = {517},
  adsurl  = {https://ui.adsabs.harvard.edu/abs/2022MNRAS.517.5104C},
  doi     = {10.1093/mnras/stac2874},
  eprint  = {2207.05771},
}

@Article{SimonDiaz+2008,
  author  = {{Sim{\'o}n-D{\'\i}az}, S. and {Stasi{\'n}ska}, G.},
  journal = {\mnras},
  title   = {{The ionizing radiation from massive stars and its impact on HII regions: results from modern model atmospheres}},
  year    = {2008},
  month   = sep,
  number  = {3},
  pages   = {1009-1021},
  volume  = {389},
  adsurl  = {https://ui.adsabs.harvard.edu/abs/2008MNRAS.389.1009S},
  doi     = {10.1111/j.1365-2966.2008.13444.x},
  eprint  = {0805.1362},
}

@Article{Leroy+2021a,
  author  = {{Leroy}, Adam K. and {Hughes}, Annie and {Liu}, Daizhong and {Pety}, J{\'e}r{\^o}me and {Rosolowsky}, Erik and {Saito}, Toshiki and {Schinnerer}, Eva and {Schruba}, Andreas and {Usero}, Antonio and {Faesi}, Christopher M. and {Herrera}, Cinthya N. and {Chevance}, M{\'e}lanie and {Hygate}, Alexander P.~S. and {Kepley}, Amanda A. and {Koch}, Eric W. and {Querejeta}, Miguel and {Sliwa}, Kazimierz and {Will}, David and {Wilson}, Christine D. and {Anand}, Gagandeep S. and {Barnes}, Ashley and {Belfiore}, Francesco and {Be{\v{s}}li{\'c}}, Ivana and {Bigiel}, Frank and {Blanc}, Guillermo A. and {Bolatto}, Alberto D. and {Boquien}, M{\`e}d{\`e}ric and {Cao}, Yixian and {Chandar}, Rupali and {Chastenet}, J{\'e}r{\'e}my and {Chiang}, I. -Da and {Congiu}, Enrico and {Dale}, Daniel A. and {Deger}, Sinan and {den Brok}, Jakob S. and {Eibensteiner}, Cosima and {Emsellem}, Eric and {Garc{\'\i}a-Rodr{\'\i}guez}, Axel and {Glover}, Simon C.~O. and {Grasha}, Kathryn and {Groves}, Brent and {Henshaw}, Jonathan D. and {Jim{\'e}nez Donaire}, Mar{\'\i}a J. and {Kim}, Jaeyeon and {Klessen}, Ralf S. and {Kreckel}, Kathryn and {Kruijssen}, J.~M. Diederik and {Larson}, Kirsten L. and {Lee}, Janice C. and {Mayker}, Ness and {McElroy}, Rebecca and {Meidt}, Sharon E. and {Mok}, Angus and {Pan}, Hsi-An and {Puschnig}, Johannes and {Razza}, Alessandro and {S{\'a}nchez-Bl'azquez}, Patricia and {Sandstrom}, Karin M. and {Santoro}, Francesco and {Sardone}, Amy and {Scheuermann}, Fabian and {Sun}, Jiayi and {Thilker}, David A. and {Turner}, Jordan A. and {Ubeda}, Leonardo and {Utomo}, Dyas and {Watkins}, Elizabeth J. and {Williams}, Thomas G.},
  journal = {\apjs},
  title   = {{PHANGS-ALMA Data Processing and Pipeline}},
  year    = {2021},
  month   = jul,
  number  = {1},
  pages   = {19},
  volume  = {255},
  adsurl  = {https://ui.adsabs.harvard.edu/abs/2021ApJS..255...19L},
  doi     = {10.3847/1538-4365/abec80},
  eid     = {19},
  eprint  = {2104.07665},
}

@Article{Leroy+2021b,
  author  = {{Leroy}, Adam K. and {Schinnerer}, Eva and {Hughes}, Annie and {Rosolowsky}, Erik and {Pety}, J{\'e}r{\^o}me and {Schruba}, Andreas and {Usero}, Antonio and {Blanc}, Guillermo A. and {Chevance}, M{\'e}lanie and {Emsellem}, Eric and {Faesi}, Christopher M. and {Herrera}, Cinthya N. and {Liu}, Daizhong and {Meidt}, Sharon E. and {Querejeta}, Miguel and {Saito}, Toshiki and {Sandstrom}, Karin M. and {Sun}, Jiayi and {Williams}, Thomas G. and {Anand}, Gagandeep S. and {Barnes}, Ashley T. and {Behrens}, Erica A. and {Belfiore}, Francesco and {Benincasa}, Samantha M. and {Be{\v{s}}li{\'c}}, Ivana and {Bigiel}, Frank and {Bolatto}, Alberto D. and {den Brok}, Jakob S. and {Cao}, Yixian and {Chandar}, Rupali and {Chastenet}, J{\'e}r{\'e}my and {Chiang}, I. -Da and {Congiu}, Enrico and {Dale}, Daniel A. and {Deger}, Sinan and {Eibensteiner}, Cosima and {Egorov}, Oleg V. and {Garc{\'\i}a-Rodr{\'\i}guez}, Axel and {Glover}, Simon C.~O. and {Grasha}, Kathryn and {Henshaw}, Jonathan D. and {Ho}, I. -Ting and {Kepley}, Amanda A. and {Kim}, Jaeyeon and {Klessen}, Ralf S. and {Kreckel}, Kathryn and {Koch}, Eric W. and {Kruijssen}, J.~M. Diederik and {Larson}, Kirsten L. and {Lee}, Janice C. and {Lopez}, Laura A. and {Machado}, Josh and {Mayker}, Ness and {McElroy}, Rebecca and {Murphy}, Eric J. and {Ostriker}, Eve C. and {Pan}, Hsi-An and {Pessa}, Ismael and {Puschnig}, Johannes and {Razza}, Alessandro and {S{\'a}nchez-Bl{\'a}zquez}, Patricia and {Santoro}, Francesco and {Sardone}, Amy and {Scheuermann}, Fabian and {Sliwa}, Kazimierz and {Sormani}, Mattia C. and {Stuber}, Sophia K. and {Thilker}, David A. and {Turner}, Jordan A. and {Utomo}, Dyas and {Watkins}, Elizabeth J. and {Whitmore}, Bradley},
  journal = {\apjs},
  title   = {{PHANGS-ALMA: Arcsecond CO(2-1) Imaging of Nearby Star-forming Galaxies}},
  year    = {2021},
  month   = dec,
  number  = {2},
  pages   = {43},
  volume  = {257},
  adsurl  = {https://ui.adsabs.harvard.edu/abs/2021ApJS..257...43L},
  doi     = {10.3847/1538-4365/ac17f3},
  eid     = {43},
  eprint  = {2104.07739},
}

@Article{Dale+2013,
  author  = {{Dale}, J.~E. and {Ercolano}, B. and {Bonnell}, I.~A.},
  journal = {\mnras},
  title   = {{Ionizing feedback from massive stars in massive clusters - III. Disruption of partially unbound clouds}},
  year    = {2013},
  month   = mar,
  number  = {1},
  pages   = {234-246},
  volume  = {430},
  adsurl  = {https://ui.adsabs.harvard.edu/abs/2013MNRAS.430..234D},
  doi     = {10.1093/mnras/sts592},
  eprint  = {1212.2011},
}

@Article{Dale+2012,
  author  = {{Dale}, J.~E. and {Ercolano}, B. and {Bonnell}, I.~A.},
  journal = {\mnras},
  title   = {{Ionizing feedback from massive stars in massive clusters - II. Disruption of bound clusters by photoionization}},
  year    = {2012},
  month   = jul,
  number  = {1},
  pages   = {377-392},
  volume  = {424},
  adsurl  = {https://ui.adsabs.harvard.edu/abs/2012MNRAS.424..377D},
  doi     = {10.1111/j.1365-2966.2012.21205.x},
  eprint  = {1205.0360},
}

@Article{Crowther+2007,
  author  = {{Crowther}, Paul A.},
  journal = {\araa},
  title   = {{Physical Properties of Wolf-Rayet Stars}},
  year    = {2007},
  month   = sep,
  number  = {1},
  pages   = {177-219},
  volume  = {45},
  adsurl  = {https://ui.adsabs.harvard.edu/abs/2007ARA&A..45..177C},
  doi     = {10.1146/annurev.astro.45.051806.110615},
  eprint  = {astro-ph/0610356},
}

@Article{Lejeune+1997,
  author  = {{Lejeune}, Th. and {Cuisinier}, F. and {Buser}, R.},
  journal = {\aaps},
  title   = {{Standard stellar library for evolutionary synthesis. I. Calibration of theoretical spectra}},
  year    = {1997},
  month   = oct,
  pages   = {229-246},
  volume  = {125},
  adsurl  = {https://ui.adsabs.harvard.edu/abs/1997A&AS..125..229L},
  doi     = {10.1051/aas:1997373},
  eprint  = {astro-ph/9701019},
}

@Article{Eldridge+2009,
  author  = {{Eldridge}, John J. and {Stanway}, Elizabeth R.},
  journal = {\mnras},
  title   = {{Spectral population synthesis including massive binaries}},
  year    = {2009},
  month   = dec,
  number  = {2},
  pages   = {1019-1028},
  volume  = {400},
  adsurl  = {https://ui.adsabs.harvard.edu/abs/2009MNRAS.400.1019E},
  doi     = {10.1111/j.1365-2966.2009.15514.x},
  eprint  = {0908.1386},
}

@Article{Krumholz+2019,
  author  = {{Krumholz}, Mark R. and {McKee}, Christopher F. and {Bland -Hawthorn}, Joss},
  journal = {\araa},
  title   = {{Star Clusters Across Cosmic Time}},
  year    = {2019},
  month   = aug,
  pages   = {227-303},
  volume  = {57},
  adsurl  = {https://ui.adsabs.harvard.edu/abs/2019ARA&A..57..227K},
  doi     = {10.1146/annurev-astro-091918-104430},
  eprint  = {1812.01615},
}

@Article{Cardelli+1989,
  author  = {{Cardelli}, Jason A. and {Clayton}, Geoffrey C. and {Mathis}, John S.},
  journal = {\apj},
  title   = {{The Relationship between Infrared, Optical, and Ultraviolet Extinction}},
  year    = {1989},
  month   = oct,
  pages   = {245},
  volume  = {345},
  adsurl  = {https://ui.adsabs.harvard.edu/abs/1989ApJ...345..245C},
  doi     = {10.1086/167900},
}

@Article{Congiu+2023,
  author  = {{Congiu}, Enrico and {Blanc}, Guillermo A. and {Belfiore}, Francesco and {Santoro}, Francesco and {Scheuermann}, Fabian and {Kreckel}, Kathryn and {Emsellem}, Eric and {Groves}, Brent and {Pan}, Hsi-An and {Bigiel}, Frank and {Dale}, Daniel A. and {Glover}, Simon C.~O. and {Grasha}, Kathryn and {Egorov}, Oleg V. and {Leroy}, Adam and {Schinnerer}, Eva and {Watkins}, Elizabeth J. and {Williams}, Thomas G.},
  journal = {\aap},
  title   = {{PHANGS-MUSE: Detection and Bayesian classification of 40 000 ionised nebulae in nearby spiral galaxies}},
  year    = {2023},
  month   = apr,
  pages   = {A148},
  volume  = {672},
  adsurl  = {https://ui.adsabs.harvard.edu/abs/2023A&A...672A.148C},
  doi     = {10.1051/0004-6361/202245153},
  eid     = {A148},
  eprint  = {2302.03062},
}

@Article{Krumholz+2015,
  author  = {{Krumholz}, Mark R. and {Adamo}, Angela and {Fumagalli}, Michele and {Wofford}, Aida and {Calzetti}, Daniela and {Lee}, Janice C. and {Whitmore}, Bradley C. and {Bright}, Stacey N. and {Grasha}, Kathryn and {Gouliermis}, Dimitrios A. and {Kim}, Hwihyun and {Nair}, Preethi and {Ryon}, Jenna E. and {Smith}, Linda J. and {Thilker}, David and {Ubeda}, Leonardo and {Zackrisson}, Erik},
  journal = {\apj},
  title   = {{Star Cluster Properties in Two LEGUS Galaxies Computed with Stochastic Stellar Population Synthesis Models}},
  year    = {2015},
  month   = oct,
  number  = {2},
  pages   = {147},
  volume  = {812},
  adsurl  = {https://ui.adsabs.harvard.edu/abs/2015ApJ...812..147K},
  doi     = {10.1088/0004-637X/812/2/147},
  eid     = {147},
  eprint  = {1509.05078},
}

@Article{Santoro+2022,
  author  = {{Santoro}, Francesco and {Kreckel}, Kathryn and {Belfiore}, Francesco and {Groves}, Brent and {Congiu}, Enrico and {Thilker}, David A. and {Blanc}, Guillermo A. and {Schinnerer}, Eva and {Ho}, I. -Ting and {Diederik Kruijssen}, J.~M. and {Meidt}, Sharon and {Klessen}, Ralf S. and {Schruba}, Andreas and {Querejeta}, Miguel and {Pessa}, Ismael and {Chevance}, M{\'e}lanie and {Kim}, Jaeyeon and {Emsellem}, Eric and {McElroy}, Rebecca and {Barnes}, Ashley T. and {Bigiel}, Frank and {Boquien}, M{\'e}d{\'e}ric and {Dale}, Daniel A. and {Glover}, Simon C.~O. and {Grasha}, Kathryn and {Lee}, Janice and {Leroy}, Adam K. and {Pan}, Hsi-An and {Rosolowsky}, Erik and {Saito}, Toshiki and {Sanchez-Blazquez}, Patricia and {Watkins}, Elizabeth J. and {Williams}, Thomas G.},
  journal = {\aap},
  title   = {{PHANGS-MUSE: The H II region luminosity function of local star-forming galaxies}},
  year    = {2022},
  month   = feb,
  pages   = {A188},
  volume  = {658},
  adsurl  = {https://ui.adsabs.harvard.edu/abs/2022A&A...658A.188S},
  doi     = {10.1051/0004-6361/202141907},
  eid     = {A188},
  eprint  = {2111.09362},
}

@Article{Thilker+2025,
  author  = {{Thilker}, David A. and {Lee}, Janice C. and {Whitmore}, Bradley C. and {Maschmann}, Daniel and {Henny}, Kiana and {Chandar}, Rupali and {Dale}, Daniel A. and {Deger}, Sinan and {Boquien}, M{\'e}d{\'e}ric and {Wofford}, Aida and {{\'U}beda}, Leonardo and {Razza}, Alessandro and {Barnes}, Ashley T. and {Belfiore}, Francesco and {Bigiel}, Frank and {Grasha}, Kathryn and {Groves}, Brent and {Kim}, Hwihyun and {Klessen}, Ralf S. and {Neumann}, Justus and {Pinna}, Francesca and {Rodr{\'\i}guez}, M. Jimena and {Rosolowsky}, Erik and {Schinnerer}, Eva and {Williams}, Thomas G.},
  journal = {\apjs},
  title   = {{PHANGS-HST Catalogs for {\ensuremath{\sim}}100,000 Star Clusters and Compact Associations in 38 Galaxies. II. Physical Properties from Decision-tree-based Spectral Energy Distribution Fitting of NUV-U-B-V-I Photometry with Categorical Priors Set by H{\ensuremath{\alpha}} Emission, Cluster Morphology, and Other Auxiliary Information}},
  year    = {2025},
  month   = sep,
  number  = {1},
  pages   = {1},
  volume  = {280},
  adsurl  = {https://ui.adsabs.harvard.edu/abs/2025ApJS..280....1T},
  doi     = {10.3847/1538-4365/addabb},
  eid     = {1},
}

@Article{Thilker+2022,
  author  = {{Thilker}, David A. and {Whitmore}, Bradley C. and {Lee}, Janice C. and {Deger}, Sinan and {Chandar}, Rupali and {Larson}, Kirsten L. and {Hannon}, Stephen and {Ubeda}, Leonardo and {Dale}, Daniel A. and {Glover}, Simon C.~O. and {Grasha}, Kathryn and {Klessen}, Ralf S. and {Kruijssen}, J.~M. Diederik and {Rosolowsky}, Erik and {Schruba}, Andreas and {White}, Richard L. and {Williams}, Thomas G.},
  journal = {\mnras},
  title   = {{PHANGS-HST: new methods for star cluster identification in nearby galaxies}},
  year    = {2022},
  month   = jan,
  number  = {3},
  pages   = {4094-4127},
  volume  = {509},
  adsurl  = {https://ui.adsabs.harvard.edu/abs/2022MNRAS.509.4094T},
  doi     = {10.1093/mnras/stab3183},
  eprint  = {2106.13366},
}

@Article{Stetson+1987,
  author  = {{Stetson}, Peter B.},
  journal = {\pasp},
  title   = {{DAOPHOT: A Computer Program for Crowded-Field Stellar Photometry}},
  year    = {1987},
  month   = mar,
  pages   = {191},
  volume  = {99},
  adsurl  = {https://ui.adsabs.harvard.edu/abs/1987PASP...99..191S},
  doi     = {10.1086/131977},
}

@Article{Cervino+2006,
  author  = {{Cervi{\~n}o}, M. and {Luridiana}, V.},
  journal = {\aap},
  title   = {{Confidence limits of evolutionary synthesis models. IV. Moving forward to a probabilistic formulation}},
  year    = {2006},
  month   = may,
  number  = {2},
  pages   = {475-498},
  volume  = {451},
  adsurl  = {https://ui.adsabs.harvard.edu/abs/2006A&A...451..475C},
  doi     = {10.1051/0004-6361:20053283},
  eprint  = {astro-ph/0504483},
}

@Article{DellaBruna+2022b,
  author  = {{Della Bruna}, Lorenza and {Adamo}, Angela and {McLeod}, Anna F. and {Smith}, Linda J. and {Savard}, Gabriel and {Robert}, Carmelle and {Sun}, Jiayi and {Amram}, Philippe and {Bik}, Arjan and {Blair}, William P. and {Long}, Knox S. and {Renaud}, Florent and {Walterbos}, Rene and {Usher}, Christopher},
  journal = {\aap},
  title   = {{Stellar feedback in M 83 as observed with MUSE. II. Analysis of the H II region population: Ionisation budget and pre-SN feedback}},
  year    = {2022},
  month   = oct,
  pages   = {A29},
  volume  = {666},
  adsurl  = {https://ui.adsabs.harvard.edu/abs/2022A&A...666A..29D},
  doi     = {10.1051/0004-6361/202243395},
  eid     = {A29},
  eprint  = {2206.09741},
}

@Article{Storey+1995,
  author  = {{Storey}, P.~J. and {Hummer}, D.~G.},
  journal = {\mnras},
  title   = {{Recombination line intensities for hydrogenic ions-IV. Total recombination coefficients and machine-readable tables for Z=1 to 8}},
  year    = {1995},
  month   = jan,
  number  = {1},
  pages   = {41-48},
  volume  = {272},
  adsurl  = {https://ui.adsabs.harvard.edu/abs/1995MNRAS.272...41S},
  doi     = {10.1093/mnras/272.1.41},
}

@Article{Dolphin+2000,
  author  = {{Dolphin}, Andrew E.},
  journal = {\pasp},
  title   = {{WFPC2 Stellar Photometry with HSTPHOT}},
  year    = {2000},
  month   = oct,
  number  = {776},
  pages   = {1383-1396},
  volume  = {112},
  adsurl  = {https://ui.adsabs.harvard.edu/abs/2000PASP..112.1383D},
  doi     = {10.1086/316630},
  eprint  = {astro-ph/0006217},
}

@Article{Kroupa+2001,
  author  = {{Kroupa}, P.},
  journal = {\mnras},
  title   = {{On the variation of the initial mass function}},
  year    = {2001},
  month   = apr,
  pages   = {231-246},
  volume  = {322},
  adsurl  = {http://adsabs.harvard.edu/abs/2001MNRAS.322..231K},
  doi     = {10.1046/j.1365-8711.2001.04022.x},
  eprint  = {astro-ph/0009005},
}

@Article{Flury+2022a,
  author  = {{Flury}, Sophia R. and {Jaskot}, Anne E. and {Ferguson}, Harry C. and {Worseck}, G{\'a}bor and {Makan}, Kirill and {Chisholm}, John and {Saldana-Lopez}, Alberto and {Schaerer}, Daniel and {McCandliss}, Stephan and {Wang}, Bingjie and {Ford}, N.~M. and {Heckman}, Timothy and {Ji}, Zhiyuan and {Giavalisco}, Mauro and {Amorin}, Ricardo and {Atek}, Hakim and {Blaizot}, Jeremy and {Borthakur}, Sanchayeeta and {Carr}, Cody and {Castellano}, Marco and {Cristiani}, Stefano and {De Barros}, Stephane and {Dickinson}, Mark and {Finkelstein}, Steven L. and {Fleming}, Brian and {Fontanot}, Fabio and {Garel}, Thibault and {Grazian}, Andrea and {Hayes}, Matthew and {Henry}, Alaina and {Mauerhofer}, Valentin and {Micheva}, Genoveva and {Oey}, M.~S. and {Ostlin}, Goran and {Papovich}, Casey and {Pentericci}, Laura and {Ravindranath}, Swara and {Rosdahl}, Joakim and {Rutkowski}, Michael and {Santini}, Paola and {Scarlata}, Claudia and {Teplitz}, Harry and {Thuan}, Trinh and {Trebitsch}, Maxime and {Vanzella}, Eros and {Verhamme}, Anne and {Xu}, Xinfeng},
  journal = {\apjs},
  title   = {{The Low-redshift Lyman Continuum Survey. I. New, Diverse Local Lyman Continuum Emitters}},
  year    = {2022},
  month   = may,
  number  = {1},
  pages   = {1},
  volume  = {260},
  adsurl  = {https://ui.adsabs.harvard.edu/abs/2022ApJS..260....1F},
  doi     = {10.3847/1538-4365/ac5331},
  eid     = {1},
  eprint  = {2201.11716},
}

@Article{DellaBruna+2021,
  author  = {{Della Bruna}, Lorenza and {Adamo}, Angela and {Lee}, Janice C. and {Smith}, Linda J. and {Krumholz}, Mark and {Bik}, Arjan and {Calzetti}, Daniela and {Fox}, Anne and {Fumagalli}, Michele and {Grasha}, Kathryn and {Messa}, Matteo and {{\"O}stlin}, G{\"o}ran and {Walterbos}, Rene and {Wofford}, Aida},
  journal = {\aap},
  title   = {{Studying the ISM at {\ensuremath{\sim}}10 pc scale in NGC 7793 with MUSE. II. Constraints on the oxygen abundance and ionising radiation escape}},
  year    = {2021},
  month   = jun,
  pages   = {A103},
  volume  = {650},
  adsurl  = {https://ui.adsabs.harvard.edu/abs/2021A&A...650A.103D},
  doi     = {10.1051/0004-6361/202039402},
  eid     = {A103},
  eprint  = {2104.08088},
}

@Article{Flury+2022b,
  author  = {{Flury}, Sophia R. and {Jaskot}, Anne E. and {Ferguson}, Harry C. and {Worseck}, G{\'a}bor and {Makan}, Kirill and {Chisholm}, John and {Saldana-Lopez}, Alberto and {Schaerer}, Daniel and {McCandliss}, Stephan R. and {Xu}, Xinfeng and {Wang}, Bingjie and {Oey}, M.~S. and {Ford}, N.~M. and {Heckman}, Timothy and {Ji}, Zhiyuan and {Giavalisco}, Mauro and {Amor{\'\i}n}, Ricardo and {Atek}, Hakim and {Blaizot}, Jeremy and {Borthakur}, Sanchayeeta and {Carr}, Cody and {Castellano}, Marco and {De Barros}, Stephane and {Dickinson}, Mark and {Finkelstein}, Steven L. and {Fleming}, Brian and {Fontanot}, Fabio and {Garel}, Thibault and {Grazian}, Andrea and {Hayes}, Matthew and {Henry}, Alaina and {Mauerhofer}, Valentin and {Micheva}, Genoveva and {Ostlin}, Goran and {Papovich}, Casey and {Pentericci}, Laura and {Ravindranath}, Swara and {Rosdahl}, Joakim and {Rutkowski}, Michael and {Santini}, Paola and {Scarlata}, Claudia and {Teplitz}, Harry and {Thuan}, Trinh and {Trebitsch}, Maxime and {Vanzella}, Eros and {Verhamme}, Anne},
  journal = {\apj},
  title   = {{The Low-redshift Lyman Continuum Survey. II. New Insights into LyC Diagnostics}},
  year    = {2022},
  month   = may,
  number  = {2},
  pages   = {126},
  volume  = {930},
  adsurl  = {https://ui.adsabs.harvard.edu/abs/2022ApJ...930..126F},
  doi     = {10.3847/1538-4357/ac61e4},
  eid     = {126},
  eprint  = {2203.15649},
}

@Article{Hollyhead+2015,
  author  = {{Hollyhead}, K. and {Bastian}, N. and {Adamo}, A. and {Silva-Villa}, E. and {Dale}, J. and {Ryon}, J.~E. and {Gazak}, Z.},
  journal = {\mnras},
  title   = {{Studying the YMC population of M83: how long clusters remain embedded, their interaction with the ISM and implications for GC formation theories}},
  year    = {2015},
  month   = may,
  number  = {1},
  pages   = {1106-1117},
  volume  = {449},
  adsurl  = {https://ui.adsabs.harvard.edu/abs/2015MNRAS.449.1106H},
  doi     = {10.1093/mnras/stv331},
  eprint  = {1502.03823},
}

@Article{RousseauNepton+2019,
  author  = {{Rousseau-Nepton}, L. and {Martin}, R.~P. and {Robert}, C. and {Drissen}, L. and {Amram}, P. and {Prunet}, S. and {Martin}, T. and {Moumen}, I. and {Adamo}, A. and {Alarie}, A. and {Barmby}, P. and {Boselli}, A. and {Bresolin}, F. and {Bureau}, M. and {Chemin}, L. and {Fernandes}, R.~C. and {Combes}, F. and {Crowder}, C. and {Della Bruna}, L. and {Duarte Puertas}, S. and {Egusa}, F. and {Epinat}, B. and {Ksoll}, V.~F. and {Girard}, M. and {G{\'o}mez Llanos}, V. and {Gouliermis}, D. and {Grasha}, K. and {Higgs}, C. and {Hlavacek-Larrondo}, J. and {Ho}, I. -T. and {Iglesias-P{\'a}ramo}, J. and {Joncas}, G. and {Kam}, Z.~S. and {Karera}, P. and {Kennicutt}, R.~C. and {Klessen}, R.~S. and {Lianou}, S. and {Liu}, L. and {Liu}, Q. and {de Amorim}, A. Luiz and {Lyman}, J.~D. and {Martel}, H. and {Mazzilli-Ciraulo}, B. and {McLeod}, A.~F. and {Melchior}, A. -L. and {Millan}, I. and {Moll{\'a}}, M. and {Momose}, R. and {Morisset}, C. and {Pan}, H. -A. and {Pati}, A.~K. and {Pellerin}, A. and {Pellegrini}, E. and {P{\'e}rez}, I. and {Petric}, A. and {Plana}, H. and {Rahner}, D. and {Ruiz Lara}, T. and {S{\'a}nchez-Menguiano}, L. and {Spekkens}, K. and {Stasi{\'n}ska}, G. and {Takamiya}, M. and {Vale Asari}, N. and {V{\'\i}lchez}, J.~M.},
  journal = {\mnras},
  title   = {{SIGNALS: I. Survey description}},
  year    = {2019},
  month   = nov,
  number  = {4},
  pages   = {5530-5546},
  volume  = {489},
  adsurl  = {https://ui.adsabs.harvard.edu/abs/2019MNRAS.489.5530R},
  doi     = {10.1093/mnras/stz2455},
  eprint  = {1908.09017},
}

@Article{Mitra+2013,
  author  = {{Mitra}, Sourav and {Ferrara}, Andrea and {Choudhury}, T. Roy},
  journal = {\mnras},
  title   = {{The escape fraction of ionizing photons from high-redshift galaxies from data-constrained reionization models}},
  year    = {2013},
  month   = jan,
  number  = {1},
  pages   = {L1-L5},
  volume  = {428},
  adsurl  = {https://ui.adsabs.harvard.edu/abs/2013MNRAS.428L...1M},
  doi     = {10.1093/mnrasl/sls001},
  eprint  = {1207.3803},
}

@Article{RousseauNepton+2018,
  author  = {{Rousseau-Nepton}, L. and {Robert}, C. and {Martin}, R.~P. and {Drissen}, L. and {Martin}, T.},
  journal = {\mnras},
  title   = {{NGC628 with SITELLE: I. Imaging spectroscopy of 4285 H II region candidates}},
  year    = {2018},
  month   = feb,
  number  = {3},
  pages   = {4152-4186},
  volume  = {477},
  adsurl  = {https://ui.adsabs.harvard.edu/abs/2018MNRAS.477.4152R},
  doi     = {10.1093/mnras/sty477},
  eprint  = {1704.05121},
}

@Article{Leitherer+2014,
  author  = {{Leitherer}, C. and {Ekstr{\"o}m}, S. and {Meynet}, G. and {Schaerer}, D. and {Agienko}, K.~B. and {Levesque}, E.~M.},
  journal = {\apjs},
  title   = {{The Effects of Stellar Rotation. II. A Comprehensive Set of Starburst99 Models}},
  year    = {2014},
  month   = May,
  pages   = {14},
  volume  = {212},
  adsurl  = {https://ui.adsabs.harvard.edu/#abs/2014ApJS..212...14L},
  doi     = {10.1088/0067-0049/212/1/14},
  eid     = {14},
  eprint  = {1403.5444},
}

@Article{Ramambason+2022,
  author  = {{Ramambason}, L. and {Lebouteiller}, V. and {Bik}, A. and {Richardson}, C.~T. and {Galliano}, F. and {Schaerer}, D. and {Morisset}, C. and {Polles}, F.~L. and {Madden}, S.~C. and {Chevance}, M. and {De Looze}, I.},
  journal = {\aap},
  title   = {{Inferring the HII region escape fraction of ionizing photons from infrared emission lines in metal-poor star-forming dwarf galaxies}},
  year    = {2022},
  month   = nov,
  pages   = {A35},
  volume  = {667},
  adsurl  = {https://ui.adsabs.harvard.edu/abs/2022A&A...667A..35R},
  doi     = {10.1051/0004-6361/202243866},
  eid     = {A35},
  eprint  = {2207.06146},
}

@Article{Leitherer+2010,
  author  = {{Leitherer}, Claus and {Ortiz Ot{\'a}lvaro}, Paula A. and {Bresolin}, Fabio and {Kudritzki}, Rolf-Peter and {Lo Faro}, Barbara and {Pauldrach}, Adalbert W.~A. and {Pettini}, Max and {Rix}, Samantha A.},
  journal = {\apjs},
  title   = {{A Library of Theoretical Ultraviolet Spectra of Massive, Hot Stars for Evolutionary Synthesis}},
  year    = {2010},
  month   = aug,
  number  = {2},
  pages   = {309-335},
  volume  = {189},
  adsurl  = {https://ui.adsabs.harvard.edu/abs/2010ApJS..189..309L},
  doi     = {10.1088/0067-0049/189/2/309},
  eprint  = {1006.5624},
}

@Article{Hannon+2022,
  author  = {{Hannon}, Stephen and {Lee}, Janice C. and {Whitmore}, B.~C. and {Mobasher}, B. and {Thilker}, D. and {Chandar}, R. and {Adamo}, A. and {Wofford}, A. and {Orozco-Duarte}, R. and {Calzetti}, D. and {Della Bruna}, L. and {Kreckel}, K. and {Groves}, B. and {Barnes}, A.~T. and {Boquien}, M. and {Belfiore}, F. and {Linden}, S.},
  journal = {\mnras},
  title   = {{H {\ensuremath{\alpha}} morphologies of star clusters in 16 LEGUS galaxies: Constraints on H II region evolution time-scales}},
  year    = {2022},
  month   = may,
  number  = {1},
  pages   = {1294-1316},
  volume  = {512},
  adsurl  = {https://ui.adsabs.harvard.edu/abs/2022MNRAS.512.1294H},
  doi     = {10.1093/mnras/stac550},
  eprint  = {2203.01339},
}

@Article{Hannon+2023,
  author  = {{Hannon}, Stephen and {Whitmore}, Bradley C. and {Lee}, Janice C. and {Thilker}, David A. and {Deger}, Sinan and {Huerta}, E.~A. and {Wei}, Wei and {Mobasher}, Bahram and {Klessen}, Ralf and {Boquien}, M{\'e}d{\'e}ric and {Dale}, Daniel A. and {Chevance}, M{\'e}lanie and {Grasha}, Kathryn and {Sanchez-Blazquez}, Patricia and {Williams}, Thomas and {Scheuermann}, Fabian and {Groves}, Brent and {Kim}, Hwihyun and {Kruijssen}, J.~M. Diederik and {The Phangs-HST Team}},
  journal = {\mnras},
  title   = {{Star cluster classification using deep transfer learning with PHANGS-HST}},
  year    = {2023},
  month   = dec,
  number  = {2},
  pages   = {2991-3006},
  volume  = {526},
  adsurl  = {https://ui.adsabs.harvard.edu/abs/2023MNRAS.526.2991H},
  doi     = {10.1093/mnras/stad2238},
  eprint  = {2307.15133},
}

@Article{Lee+2022,
  author  = {{Lee}, Janice C. and {Whitmore}, Bradley C. and {Thilker}, David A. and {Deger}, Sinan and {Larson}, Kirsten L. and {Ubeda}, Leonardo and {Anand}, Gagandeep S. and {Boquien}, M{\'e}d{\'e}ric and {Chandar}, Rupali and {Dale}, Daniel A. and {Emsellem}, Eric and {Leroy}, Adam K. and {Rosolowsky}, Erik and {Schinnerer}, Eva and {Schmidt}, Judy and {Lilly}, James and {Turner}, Jordan and {Van Dyk}, Schuyler and {White}, Richard L. and {Barnes}, Ashley T. and {Belfiore}, Francesco and {Bigiel}, Frank and {Blanc}, Guillermo A. and {Cao}, Yixian and {Chevance}, Melanie and {Congiu}, Enrico and {Egorov}, Oleg V. and {Glover}, Simon C.~O. and {Grasha}, Kathryn and {Groves}, Brent and {Henshaw}, Jonathan D. and {Hughes}, Annie and {Klessen}, Ralf S. and {Koch}, Eric and {Kreckel}, Kathryn and {Kruijssen}, J.~M. Diederik and {Liu}, Daizhong and {Lopez}, Laura A. and {Mayker}, Ness and {Meidt}, Sharon E. and {Murphy}, Eric J. and {Pan}, Hsi-An and {Pety}, J{\'e}r{\^o}me and {Querejeta}, Miguel and {Razza}, Alessandro and {Saito}, Toshiki and {S{\'a}nchez-Bl{\'a}zquez}, Patricia and {Santoro}, Francesco and {Sardone}, Amy and {Scheuermann}, Fabian and {Schruba}, Andreas and {Sun}, Jiayi and {Usero}, Antonio and {Watkins}, E. and {Williams}, Thomas G.},
  journal = {\apjs},
  title   = {{The PHANGS-HST Survey: Physics at High Angular Resolution in Nearby Galaxies with the Hubble Space Telescope}},
  year    = {2022},
  month   = jan,
  number  = {1},
  pages   = {10},
  volume  = {258},
  adsurl  = {https://ui.adsabs.harvard.edu/abs/2022ApJS..258...10L},
  doi     = {10.3847/1538-4365/ac1fe5},
  eid     = {10},
  eprint  = {2101.02855},
}

@Article{Lee+2023,
  author  = {{Lee}, Janice C. and {Sandstrom}, Karin M. and {Leroy}, Adam K. and {Thilker}, David A. and {Schinnerer}, Eva and {Rosolowsky}, Erik and {Larson}, Kirsten L. and {Egorov}, Oleg V. and {Williams}, Thomas G. and {Schmidt}, Judy and {Emsellem}, Eric and {Anand}, Gagandeep S. and {Barnes}, Ashley T. and {Belfiore}, Francesco and {Be{\v{s}}li{\'c}}, Ivana and {Bigiel}, Frank and {Blanc}, Guillermo A. and {Bolatto}, Alberto D. and {Boquien}, M{\'e}d{\'e}ric and {den Brok}, Jakob and {Cao}, Yixian and {Chandar}, Rupali and {Chastenet}, J{\'e}r{\'e}my and {Chevance}, M{\'e}lanie and {Chiang}, I-Da and {Congiu}, Enrico and {Dale}, Daniel A. and {Deger}, Sinan and {Eibensteiner}, Cosima and {Faesi}, Christopher M. and {Glover}, Simon C.~O. and {Grasha}, Kathryn and {Groves}, Brent and {Hassani}, Hamid and {Henny}, Kiana F. and {Henshaw}, Jonathan D. and {Hoyer}, Nils and {Hughes}, Annie and {Jeffreson}, Sarah and {Jim{\'e}nez-Donaire}, Mar{\'\i}a J. and {Kim}, Jaeyeon and {Kim}, Hwihyun and {Klessen}, Ralf S. and {Koch}, Eric W. and {Kreckel}, Kathryn and {Kruijssen}, J.~M. Diederik and {Li}, Jing and {Liu}, Daizhong and {Lopez}, Laura A. and {Maschmann}, Daniel and {Chen}, Ness Mayker and {Meidt}, Sharon E. and {Murphy}, Eric J. and {Neumann}, Justus and {Neumayer}, Nadine and {Pan}, Hsi-An and {Pessa}, Ismael and {Pety}, J{\'e}r{\^o}me and {Querejeta}, Miguel and {Pinna}, Francesca and {Rodr{\'\i}guez}, M. Jimena and {Saito}, Toshiki and {S{\'a}nchez-Bl{\'a}zquez}, Patricia and {Santoro}, Francesco and {Sardone}, Amy and {Smith}, Rowan J. and {Sormani}, Mattia C. and {Scheuermann}, Fabian and {Stuber}, Sophia K. and {Sutter}, Jessica and {Sun}, Jiayi and {Teng}, Yu-Hsuan and {Tre{\ss}}, Robin G. and {Usero}, Antonio and {Watkins}, Elizabeth J. and {Whitmore}, Bradley C. and {Razza}, Alessandro},
  journal = {\apjl},
  title   = {{The PHANGS-JWST Treasury Survey: Star Formation, Feedback, and Dust Physics at High Angular Resolution in Nearby GalaxieS}},
  year    = {2023},
  month   = feb,
  number  = {2},
  pages   = {L17},
  volume  = {944},
  adsurl  = {https://ui.adsabs.harvard.edu/abs/2023ApJ...944L..17L},
  doi     = {10.3847/2041-8213/acaaae},
  eid     = {L17},
  eprint  = {2212.02667},
}

@Article{Ramambason+2020,
  author  = {{Ramambason}, L. and {Schaerer}, D. and {Stasi{\'n}ska}, G. and {Izotov}, Y.~I. and {Guseva}, N.~G. and {V{\'\i}lchez}, J.~M. and {Amor{\'\i}n}, R. and {Morisset}, C.},
  journal = {\aap},
  title   = {{Reconciling escape fractions and observed line emission in Lyman-continuum-leaking galaxies}},
  year    = {2020},
  month   = dec,
  pages   = {A21},
  volume  = {644},
  adsurl  = {https://ui.adsabs.harvard.edu/abs/2020A&A...644A..21R},
  doi     = {10.1051/0004-6361/202038634},
  eid     = {A21},
  eprint  = {2009.09882},
}

@Article{Ma+2020,
  author  = {{Ma}, Xiangcheng and {Quataert}, Eliot and {Wetzel}, Andrew and {Hopkins}, Philip F. and {Faucher-Gigu{\`e}re}, Claude-Andr{\'e} and {Kere{\v{s}}}, Du{\v{s}}an},
  journal = {\mnras},
  title   = {{No missing photons for reionization: moderate ionizing photon escape fractions from the FIRE-2 simulations}},
  year    = {2020},
  month   = oct,
  number  = {2},
  pages   = {2001-2017},
  volume  = {498},
  adsurl  = {https://ui.adsabs.harvard.edu/abs/2020MNRAS.498.2001M},
  doi     = {10.1093/mnras/staa2404},
  eprint  = {2003.05945},
}

@InProceedings{Bacon+2010,
  author    = {{Bacon}, R. and {Accardo}, M. and {Adjali}, L. and {Anwand}, H. and {Bauer}, S. and {Biswas}, I. and {Blaizot}, J. and {Boudon}, D. and {Brau-Nogue}, S. and {Brinchmann}, J. and {Caillier}, P. and {Capoani}, L. and {Carollo}, C.~M. and {Contini}, T. and {Couderc}, P. and {Daguis{\'e}}, E. and {Deiries}, S. and {Delabre}, B. and {Dreizler}, S. and {Dubois}, J. and {Dupieux}, M. and {Dupuy}, C. and {Emsellem}, E. and {Fechner}, T. and {Fleischmann}, A. and {Fran{\c{c}}ois}, M. and {Gallou}, G. and {Gharsa}, T. and {Glindemann}, A. and {Gojak}, D. and {Guiderdoni}, B. and {Hansali}, G. and {Hahn}, T. and {Jarno}, A. and {Kelz}, A. and {Koehler}, C. and {Kosmalski}, J. and {Laurent}, F. and {Le Floch}, M. and {Lilly}, S.~J. and {Lizon}, J. -L. and {Loupias}, M. and {Manescau}, A. and {Monstein}, C. and {Nicklas}, H. and {Olaya}, J. -C. and {Pares}, L. and {Pasquini}, L. and {P{\'e}contal-Rousset}, A. and {Pell{\'o}}, R. and {Petit}, C. and {Popow}, E. and {Reiss}, R. and {Remillieux}, A. and {Renault}, E. and {Roth}, M. and {Rupprecht}, G. and {Serre}, D. and {Schaye}, J. and {Soucail}, G. and {Steinmetz}, M. and {Streicher}, O. and {Stuik}, R. and {Valentin}, H. and {Vernet}, J. and {Weilbacher}, P. and {Wisotzki}, L. and {Yerle}, N.},
  booktitle = {Ground-based and Airborne Instrumentation for Astronomy III},
  title     = {{The MUSE second-generation VLT instrument}},
  year      = {2010},
  month     = jul,
  pages     = {773508},
  series    = {\procspie},
  volume    = {7735},
  adsurl    = {http://adsabs.harvard.edu/abs/2010SPIE.7735E..08B},
  doi       = {10.1117/12.856027},
  eid       = {773508},
}

@Article{Fouesneau+2012,
  author  = {{Fouesneau}, Morgan and {Lan{\c{c}}on}, Ariane and {Chandar}, Rupali and {Whitmore}, Bradley C.},
  journal = {\apj},
  title   = {{Analyzing Star Cluster Populations with Stochastic Models: The Hubble Space Telescope/Wide Field Camera 3 Sample of Clusters in M83}},
  year    = {2012},
  month   = may,
  number  = {1},
  pages   = {60},
  volume  = {750},
  adsurl  = {https://ui.adsabs.harvard.edu/abs/2012ApJ...750...60F},
  doi     = {10.1088/0004-637X/750/1/60},
  eid     = {60},
  eprint  = {1202.3135},
}

@Article{Tacchella+2022,
  author  = {{Tacchella}, Sandro and {Smith}, Aaron and {Kannan}, Rahul and {Marinacci}, Federico and {Hernquist}, Lars and {Vogelsberger}, Mark and {Torrey}, Paul and {Sales}, Laura and {Li}, Hui},
  journal = {\mnras},
  title   = {{H {\ensuremath{\alpha}} emission in local galaxies: star formation, time variability, and the diffuse ionized gas}},
  year    = {2022},
  month   = jun,
  number  = {2},
  pages   = {2904-2929},
  volume  = {513},
  adsurl  = {https://ui.adsabs.harvard.edu/abs/2022MNRAS.513.2904T},
  doi     = {10.1093/mnras/stac818},
  eprint  = {2112.00027},
}

@Article{Boquien+2019,
  author  = {{Boquien}, M. and {Burgarella}, D. and {Roehlly}, Y. and {Buat}, V. and {Ciesla}, L. and {Corre}, D. and {Inoue}, A.~K. and {Salas}, H.},
  journal = {\aap},
  title   = {{CIGALE: a python Code Investigating GALaxy Emission}},
  year    = {2019},
  month   = feb,
  pages   = {A103},
  volume  = {622},
  adsurl  = {https://ui.adsabs.harvard.edu/abs/2019A&A...622A.103B},
  doi     = {10.1051/0004-6361/201834156},
  eid     = {A103},
  eprint  = {1811.03094},
}

@Article{Gerasimov+2022,
  author  = {{Gerasimov}, Ivan S. and {Egorov}, Oleg V. and {Lozinskaya}, Tatiana A. and {Moiseev}, Alexei V. and {Oparin}, Dmitry V.},
  journal = {\mnras},
  title   = {{Stellar feedback impact on the ionized gas kinematics in the dwarf galaxy Sextans A}},
  year    = {2022},
  month   = dec,
  number  = {4},
  pages   = {4968-4985},
  volume  = {517},
  adsurl  = {https://ui.adsabs.harvard.edu/abs/2022MNRAS.517.4968G},
  doi     = {10.1093/mnras/stac3002},
  eprint  = {2210.07726},
}

@Article{Belfiore+2022,
  author  = {{Belfiore}, F. and {Santoro}, F. and {Groves}, B. and {Schinnerer}, E. and {Kreckel}, K. and {Glover}, S.~C.~O. and {Klessen}, R.~S. and {Emsellem}, E. and {Blanc}, G.~A. and {Congiu}, E. and {Barnes}, A.~T. and {Boquien}, M. and {Chevance}, M. and {Dale}, D.~A. and {Diederik Kruijssen}, J.~M. and {Leroy}, A.~K. and {Pan}, H. -A. and {Pessa}, I. and {Schruba}, A. and {Williams}, T.~G.},
  journal = {\aap},
  title   = {{A tale of two DIGs: The relative role of H II regions and low-mass hot evolved stars in powering the diffuse ionised gas (DIG) in PHANGS-MUSE galaxies}},
  year    = {2022},
  month   = mar,
  pages   = {A26},
  volume  = {659},
  adsurl  = {https://ui.adsabs.harvard.edu/abs/2022A&A...659A..26B},
  doi     = {10.1051/0004-6361/202141859},
  eid     = {A26},
  eprint  = {2111.14876},
}

@Article{Fouesneau+2010,
  author  = {{Fouesneau}, M. and {Lan{\c{c}}on}, A.},
  journal = {\aap},
  title   = {{Accounting for stochastic fluctuations when analysing the integrated light of star clusters. I. First systematics}},
  year    = {2010},
  month   = oct,
  pages   = {A22},
  volume  = {521},
  adsurl  = {https://ui.adsabs.harvard.edu/abs/2010A&A...521A..22F},
  doi     = {10.1051/0004-6361/201014084},
  eid     = {A22},
  eprint  = {1003.2334},
}

@Article{Lopez+2014,
  author  = {{Lopez}, Laura A. and {Krumholz}, Mark R. and {Bolatto}, Alberto D. and {Prochaska}, J. Xavier and {Ramirez-Ruiz}, Enrico and {Castro}, Daniel},
  journal = {\apj},
  title   = {{The Role of Stellar Feedback in the Dynamics of H II Regions}},
  year    = {2014},
  month   = nov,
  number  = {2},
  pages   = {121},
  volume  = {795},
  adsurl  = {https://ui.adsabs.harvard.edu/abs/2014ApJ...795..121L},
  doi     = {10.1088/0004-637X/795/2/121},
  eid     = {121},
  eprint  = {1309.5421},
}

@Article{Levesque+2012,
  author  = {{Levesque}, Emily M. and {Leitherer}, Claus and {Ekstrom}, Sylvia and {Meynet}, Georges and {Schaerer}, Daniel},
  journal = {\apj},
  title   = {{The Effects of Stellar Rotation. I. Impact on the Ionizing Spectra and Integrated Properties of Stellar Populations}},
  year    = {2012},
  month   = may,
  number  = {1},
  pages   = {67},
  volume  = {751},
  adsurl  = {https://ui.adsabs.harvard.edu/abs/2012ApJ...751...67L},
  doi     = {10.1088/0004-637X/751/1/67},
  eid     = {67},
  eprint  = {1203.5109},
}

@Article{Zastrow+2013,
  author  = {{Zastrow}, Jordan and {Oey}, M.~S. and {Pellegrini}, E.~W.},
  journal = {\apj},
  title   = {{Single-star H II Regions as a Probe of Massive Star Spectral Energy Distributions}},
  year    = {2013},
  month   = jun,
  number  = {2},
  pages   = {94},
  volume  = {769},
  adsurl  = {https://ui.adsabs.harvard.edu/abs/2013ApJ...769...94Z},
  doi     = {10.1088/0004-637X/769/2/94},
  eid     = {94},
  eprint  = {1212.5487},
}

@Article{Groves+2023,
  author  = {{Groves}, B. and {Kreckel}, K. and {Santoro}, F. and {Belfiore}, F. and {Zavodnik}, E. and {Congiu}, E. and {Egorov}, O.~V. and {Emsellem}, E. and {Grasha}, K. and {Leroy}, A. and {Scheuermann}, F. and {Schinnerer}, E. and {Watkins}, E.~J. and {Barnes}, A.~T. and {Bigiel}, F. and {Dale}, D.~A. and {Glover}, S.~C.~O. and {Pessa}, I. and {Sanchez-Blazquez}, P. and {Williams}, T.~G.},
  journal = {\mnras},
  title   = {{The PHANGS-MUSE nebular catalogue}},
  year    = {2023},
  month   = apr,
  number  = {4},
  pages   = {4902-4952},
  volume  = {520},
  adsurl  = {https://ui.adsabs.harvard.edu/abs/2023MNRAS.520.4902G},
  doi     = {10.1093/mnras/stad114},
  eprint  = {2301.03811},
}

@Article{Gouliermis+2018,
  author  = {{Gouliermis}, Dimitrios A.},
  journal = {\pasp},
  title   = {{Unbound Young Stellar Systems: Star Formation on the Loose}},
  year    = {2018},
  month   = jul,
  number  = {989},
  pages   = {072001},
  volume  = {130},
  adsurl  = {https://ui.adsabs.harvard.edu/abs/2018PASP..130g2001G},
  doi     = {10.1088/1538-3873/aac1fd},
  eprint  = {1806.11541},
}

@Article{Emsellem+2022,
  author  = {{Emsellem}, Eric and {Schinnerer}, Eva and {Santoro}, Francesco and {Belfiore}, Francesco and {Pessa}, Ismael and {McElroy}, Rebecca and {Blanc}, Guillermo A. and {Congiu}, Enrico and {Groves}, Brent and {Ho}, I. -Ting and {Kreckel}, Kathryn and {Razza}, Alessandro and {Sanchez-Blazquez}, Patricia and {Egorov}, Oleg and {Faesi}, Chris and {Klessen}, Ralf S. and {Leroy}, Adam K. and {Meidt}, Sharon and {Querejeta}, Miguel and {Rosolowsky}, Erik and {Scheuermann}, Fabian and {Anand}, Gagandeep S. and {Barnes}, Ashley T. and {Be{\v{s}}li{\'c}}, Ivana and {Bigiel}, Frank and {Boquien}, M{\'e}d{\'e}ric and {Cao}, Yixian and {Chevance}, M{\'e}lanie and {Dale}, Daniel A. and {Eibensteiner}, Cosima and {Glover}, Simon C.~O. and {Grasha}, Kathryn and {Henshaw}, Jonathan D. and {Hughes}, Annie and {Koch}, Eric W. and {Kruijssen}, J.~M. Diederik and {Lee}, Janice and {Liu}, Daizhong and {Pan}, Hsi-An and {Pety}, J{\'e}r{\^o}me and {Saito}, Toshiki and {Sandstrom}, Karin M. and {Schruba}, Andreas and {Sun}, Jiayi and {Thilker}, David A. and {Usero}, Antonio and {Watkins}, Elizabeth J. and {Williams}, Thomas G.},
  journal = {\aap},
  title   = {{The PHANGS-MUSE survey. Probing the chemo-dynamical evolution of disc galaxies}},
  year    = {2022},
  month   = mar,
  pages   = {A191},
  volume  = {659},
  adsurl  = {https://ui.adsabs.harvard.edu/abs/2022A&A...659A.191E},
  doi     = {10.1051/0004-6361/202141727},
  eid     = {A191},
  eprint  = {2110.03708},
}

@Article{Adamo+2017,
  author  = {{Adamo}, A. and {Ryon}, J.~E. and {Messa}, M. and {Kim}, H. and {Grasha}, K. and {Cook}, D.~O. and {Calzetti}, D. and {Lee}, J.~C. and {Whitmore}, B.~C. and {Elmegreen}, B.~G. and {Ubeda}, L. and {Smith}, L.~J. and {Bright}, S.~N. and {Runnholm}, A. and {Andrews}, J.~E. and {Fumagalli}, M. and {Gouliermis}, D.~A. and {Kahre}, L. and {Nair}, P. and {Thilker}, D. and {Walterbos}, R. and {Wofford}, A. and {Aloisi}, A. and {Ashworth}, G. and {Brown}, T.~M. and {Chandar}, R. and {Christian}, C. and {Cignoni}, M. and {Clayton}, G.~C. and {Dale}, D.~A. and {de Mink}, S.~E. and {Dobbs}, C. and {Elmegreen}, D.~M. and {Evans}, A.~S. and {Gallagher}, J.~S., III and {Grebel}, E.~K. and {Herrero}, A. and {Hunter}, D.~A. and {Johnson}, K.~E. and {Kennicutt}, R.~C. and {Krumholz}, M.~R. and {Lennon}, D. and {Levay}, K. and {Martin}, C. and {Nota}, A. and {{\"O}stlin}, G. and {Pellerin}, A. and {Prieto}, J. and {Regan}, M.~W. and {Sabbi}, E. and {Sacchi}, E. and {Schaerer}, D. and {Schiminovich}, D. and {Shabani}, F. and {Tosi}, M. and {Van Dyk}, S.~D. and {Zackrisson}, E.},
  journal = {\apj},
  title   = {{Legacy ExtraGalactic UV Survey with The Hubble Space Telescope: Stellar Cluster Catalogs and First Insights Into Cluster Formation and Evolution in NGC 628}},
  year    = {2017},
  month   = jun,
  number  = {2},
  pages   = {131},
  volume  = {841},
  adsurl  = {https://ui.adsabs.harvard.edu/abs/2017ApJ...841..131A},
  doi     = {10.3847/1538-4357/aa7132},
  eid     = {131},
  eprint  = {1705.01588},
}

@Article{Turner+2021,
  author  = {{Turner}, Jordan A. and {Dale}, Daniel A. and {Lee}, Janice C. and {Boquien}, M{\'e}d{\'e}ric and {Chandar}, Rupali and {Deger}, Sinan and {Larson}, Kirsten L. and {Mok}, Angus and {Thilker}, David A. and {Ubeda}, Leonardo and {Whitmore}, Bradley C. and {Belfiore}, Francesco and {Bigiel}, Frank and {Blanc}, Guillermo A. and {Emsellem}, Eric and {Grasha}, Kathryn and {Groves}, Brent and {Klessen}, Ralf S. and {Kreckel}, Kathryn and {Kruijssen}, J.~M. Diederik and {Leroy}, Adam K. and {Rosolowsky}, Erik and {Sanchez-Blazquez}, Patricia and {Schinnerer}, Eva and {Schruba}, Andreas and {Van Dyk}, Schuyler D. and {Williams}, Thomas G.},
  journal = {\mnras},
  title   = {{PHANGS-HST: star cluster spectral energy distribution fitting with CIGALE}},
  year    = {2021},
  month   = mar,
  number  = {1},
  pages   = {1366-1385},
  volume  = {502},
  adsurl  = {https://ui.adsabs.harvard.edu/abs/2021MNRAS.502.1366T},
  doi     = {10.1093/mnras/stab055},
  eprint  = {2101.02134},
}

@Article{KadoFong+2020,
  author  = {{Kado-Fong}, Erin and {Kim}, Jeong-Gyu and {Ostriker}, Eve C. and {Kim}, Chang-Goo},
  journal = {\apj},
  title   = {{Diffuse Ionized Gas in Simulations of Multiphase, Star-forming Galactic Disks}},
  year    = {2020},
  month   = jul,
  number  = {2},
  pages   = {143},
  volume  = {897},
  adsurl  = {https://ui.adsabs.harvard.edu/abs/2020ApJ...897..143K},
  doi     = {10.3847/1538-4357/ab9abd},
  eid     = {143},
  eprint  = {2006.06697},
}

@Article{Chandar+2025,
  author  = {{Chandar}, Rupali and {Barnes}, Ashley T. and {Thilker}, David A. and {Caputo}, Miranda and {Floyd}, Matthew R. and {Leroy}, Adam K. and {{\'U}beda}, Leonardo and {Lee}, Janice C. and {Boquien}, M{\'e}d{\'e}ric and {Maschmann}, Daniel and {Belfiore}, Francesco and {Kreckel}, Kathryn and {Glover}, Simon C.~O. and {Klessen}, Ralf S. and {Groves}, Brent and {Dale}, Daniel A. and {Schinnerer}, Eva and {Emsellem}, Eric and {Rosolowsky}, Erik and {Bigiel}, Frank and {Blanc}, Guillermo and {Chevance}, M{\'e}lanie and {Congiu}, Enrico and {Egorov}, Oleg V. and {Faesi}, Chris and {Grasha}, Kathryn and {Hannon}, Stephen and {Larson}, Kirsten L. and {Lopez}, Laura A. and {Mok}, Angus and {Neumann}, Justus and {Ostriker}, Eve and {Razza}, Alessandro and {S{\'a}nchez-Bl{\'a}zquez}, Patricia and {Santoro}, Francesco and {Schruba}, Andreas and {Sun}, Jiayi and {Usero}, Antonio and {Watkins}, E. and {Whitmore}, Bradley C. and {Williams}, Thomas G.},
  journal = {\aj},
  title   = {{The PHANGS-HST-H{\ensuremath{\alpha}} Survey: Warm Ionized Gas Physics at High Angular Resolution in Nearby Galaxies with the Hubble Space Telescope}},
  year    = {2025},
  month   = mar,
  number  = {3},
  pages   = {150},
  volume  = {169},
  adsurl  = {https://ui.adsabs.harvard.edu/abs/2025AJ....169..150C},
  doi     = {10.3847/1538-3881/adaa80},
  eid     = {150},
  eprint  = {2503.18791},
}

@Article{Hanish+2010,
  author  = {{Hanish}, D.~J. and {Oey}, M.~S. and {Rigby}, J.~R. and {de Mello}, D.~F. and {Lee}, J.~C.},
  journal = {\apj},
  title   = {{A Multiwavelength Study on the Fate of Ionizing Radiation in Local Starbursts}},
  year    = {2010},
  month   = dec,
  number  = {2},
  pages   = {2029-2037},
  volume  = {725},
  adsurl  = {https://ui.adsabs.harvard.edu/abs/2010ApJ...725.2029H},
  doi     = {10.1088/0004-637X/725/2/2029},
  eprint  = {1010.2727},
}

@Article{Jacobs+2009,
  author  = {{Jacobs}, Bradley A. and {Rizzi}, Luca and {Tully}, R. Brent and {Shaya}, Edward J. and {Makarov}, Dmitry I. and {Makarova}, Lidia},
  journal = {\aj},
  title   = {{The Extragalactic Distance Database: Color-Magnitude Diagrams}},
  year    = {2009},
  month   = aug,
  number  = {2},
  pages   = {332-337},
  volume  = {138},
  adsurl  = {https://ui.adsabs.harvard.edu/abs/2009AJ....138..332J},
  doi     = {10.1088/0004-6256/138/2/332},
  eprint  = {0902.3675},
}

@Article{daSilva+2012,
  author  = {{da Silva}, Robert L. and {Fumagalli}, Michele and {Krumholz}, Mark},
  journal = {\apj},
  title   = {{SLUG{\textemdash}Stochastically Lighting Up Galaxies. I. Methods and Validating Tests}},
  year    = {2012},
  month   = feb,
  number  = {2},
  pages   = {145},
  volume  = {745},
  adsurl  = {https://ui.adsabs.harvard.edu/abs/2012ApJ...745..145D},
  doi     = {10.1088/0004-637X/745/2/145},
  eid     = {145},
  eprint  = {1106.3072},
}

@Article{Larson+2023,
  author  = {{Larson}, Kirsten L. and {Lee}, Janice C. and {Thilker}, David A. and {Whitmore}, Bradley C. and {Deger}, Sinan and {Lilly}, James and {Chandar}, Rupali and {Dale}, Daniel A. and {Bigiel}, Frank and {Grasha}, Kathryn and {Groves}, Brent and {Hannon}, Stephen and {Klessen}, Ralf S. and {Kreckel}, Kathryn and {Kruijssen}, J.~M. Diederik and {Leroy}, Adam K. and {Pan}, Hsi-An and {Rosolowsky}, Erik and {Schinnerer}, Eva and {Schruba}, Andreas and {Watkins}, Elizabeth J. and {Williams}, Thomas G.},
  journal = {\mnras},
  title   = {{Multiscale stellar associations across the star formation hierarchy in PHANGS-HST nearby galaxies: methodology and properties}},
  year    = {2023},
  month   = aug,
  number  = {4},
  pages   = {6061-6081},
  volume  = {523},
  adsurl  = {https://ui.adsabs.harvard.edu/abs/2023MNRAS.523.6061L},
  doi     = {10.1093/mnras/stad1600},
  eprint  = {2212.11425},
}

@Article{Harris+2020,
  author    = {Charles R. Harris and K. Jarrod Millman and St{\'{e}}fan J. van der Walt and Ralf Gommers and Pauli Virtanen and David Cournapeau and Eric Wieser and Julian Taylor and Sebastian Berg and Nathaniel J. Smith and Robert Kern and Matti Picus and Stephan Hoyer and Marten H. van Kerkwijk and Matthew Brett and Allan Haldane and Jaime Fern{\'{a}}ndez del R{\'{i}}o and Mark Wiebe and Pearu Peterson and Pierre G{\'{e}}rard-Marchant and Kevin Sheppard and Tyler Reddy and Warren Weckesser and Hameer Abbasi and Christoph Gohlke and Travis E. Oliphant},
  journal   = {\nat},
  title     = {Array programming with {NumPy}},
  year      = {2020},
  month     = sep,
  number    = {7825},
  pages     = {357--362},
  volume    = {585},
  doi       = {10.1038/s41586-020-2649-2},
  publisher = {Springer Science and Business Media {LLC}},
  url       = {https://doi.org/10.1038/s41586-020-2649-2},
}

@Article{Egorov+2018,
  author  = {{Egorov}, Oleg V. and {Lozinskaya}, Tatiana A. and {Moiseev}, Alexei V. and {Smirnov-Pinchukov}, Grigorii V.},
  journal = {\mnras},
  title   = {{Star formation complexes in the `galaxy-sized' supergiant shell of the galaxy HolmbergI}},
  year    = {2018},
  month   = aug,
  number  = {3},
  pages   = {3386-3409},
  volume  = {478},
  adsurl  = {https://ui.adsabs.harvard.edu/abs/2018MNRAS.478.3386E},
  doi     = {10.1093/mnras/sty1158},
  eprint  = {1805.00315},
}

@Article{Kim+2021,
  author  = {{Kim}, Jaeyeon and {Chevance}, M{\'e}lanie and {Kruijssen}, J.~M. Diederik and {Schruba}, Andreas and {Sandstrom}, Karin and {Barnes}, Ashley T. and {Bigiel}, Frank and {Blanc}, Guillermo A. and {Cao}, Yixian and {Dale}, Daniel A. and {Faesi}, Christopher M. and {Glover}, Simon C.~O. and {Grasha}, Kathryn and {Groves}, Brent and {Herrera}, Cinthya and {Klessen}, Ralf S. and {Kreckel}, Kathryn and {Lee}, Janice C. and {Leroy}, Adam K. and {Pety}, J{\'e}r{\^o}me and {Querejeta}, Miguel and {Schinnerer}, Eva and {Sun}, Jiayi and {Usero}, Antonio and {Ward}, Jacob L. and {Williams}, Thomas G.},
  journal = {\mnras},
  title   = {{On the duration of the embedded phase of star formation}},
  year    = {2021},
  month   = jun,
  number  = {1},
  pages   = {487-509},
  volume  = {504},
  adsurl  = {https://ui.adsabs.harvard.edu/abs/2021MNRAS.504..487K},
  doi     = {10.1093/mnras/stab878},
  eprint  = {2012.00019},
}

@Article{Kim+2023,
  author  = {{Kim}, Jeong-Gyu and {Gong}, Munan and {Kim}, Chang-Goo and {Ostriker}, Eve C.},
  journal = {\apjs},
  title   = {{Photochemistry and Heating/Cooling of the Multiphase Interstellar Medium with UV Radiative Transfer for Magnetohydrodynamic Simulations}},
  year    = {2023},
  month   = jan,
  number  = {1},
  pages   = {10},
  volume  = {264},
  adsurl  = {https://ui.adsabs.harvard.edu/abs/2023ApJS..264...10K},
  doi     = {10.3847/1538-4365/ac9b1d},
  eid     = {10},
  eprint  = {2210.08024},
}

@Article{Veilleux+1987,
  author  = {{Veilleux}, Sylvain and {Osterbrock}, Donald E.},
  journal = {\apjs},
  title   = {{Spectral Classification of Emission-Line Galaxies}},
  year    = {1987},
  month   = feb,
  pages   = {295},
  volume  = {63},
  adsurl  = {https://ui.adsabs.harvard.edu/abs/1987ApJS...63..295V},
  doi     = {10.1086/191166},
}

@Article{Kim+2021a,
  author  = {{Kim}, Jeong-Gyu and {Ostriker}, Eve C. and {Filippova}, Nina},
  journal = {\apj},
  title   = {{Star Formation Efficiency and Dispersal of Giant Molecular Clouds with UV Radiation Feedback: Dependence on Gravitational Boundedness and Magnetic Fields}},
  year    = {2021},
  month   = apr,
  number  = {2},
  pages   = {128},
  volume  = {911},
  adsurl  = {https://ui.adsabs.harvard.edu/abs/2021ApJ...911..128K},
  doi     = {10.3847/1538-4357/abe934},
  eid     = {128},
  eprint  = {2011.07772},
}

@Article{Maschmann+2024,
  author  = {{Maschmann}, Daniel and {Lee}, Janice C. and {Thilker}, David A. and {Whitmore}, Bradley C. and {Deger}, Sinan and {Boquien}, M{\'e}d{\'e}ric and {Chandar}, Rupali and {Dale}, Daniel A. and {Wofford}, Aida and {Hannon}, Stephen and {Larson}, Kirsten L. and {Leroy}, Adam K. and {Schinnerer}, Eva and {Rosolowsky}, Erik and {{\'U}beda}, Leonardo and {Barnes}, Ashley T. and {Emsellem}, Eric and {Grasha}, Kathryn and {Groves}, Brent and {Indebetouw}, R{\'e}my and {Kim}, Hwihyun and {Klessen}, Ralf S. and {Kreckel}, Kathryn and {Levy}, Rebecca C. and {Pinna}, Francesca and {Rodr{\'\i}guez}, M. Jimena and {Tian}, Qiushi and {Williams}, Thomas G.},
  journal = {\apjs},
  title   = {{PHANGS-HST Catalogs for {\ensuremath{\sim}}100,000 Star Clusters and Compact Associations in 38 Galaxies. I. Observed Properties}},
  year    = {2024},
  month   = jul,
  number  = {1},
  pages   = {14},
  volume  = {273},
  adsurl  = {https://ui.adsabs.harvard.edu/abs/2024ApJS..273...14M},
  doi     = {10.3847/1538-4365/ad3cd3},
  eid     = {14},
  eprint  = {2403.04901},
}

@Article{daSilva+2014,
  author  = {{da Silva}, Robert L. and {Fumagalli}, Michele and {Krumholz}, Mark R.},
  journal = {\mnras},
  title   = {{SLUG - Stochastically Lighting Up Galaxies - II. Quantifying the effects of stochasticity on star formation rate indicators}},
  year    = {2014},
  month   = nov,
  number  = {4},
  pages   = {3275-3287},
  volume  = {444},
  adsurl  = {https://ui.adsabs.harvard.edu/abs/2014MNRAS.444.3275D},
  doi     = {10.1093/mnras/stu1688},
  eprint  = {1403.4605},
}

@Article{Dayal+2018,
  author  = {{Dayal}, Pratika and {Ferrara}, Andrea},
  journal = {\physrep},
  title   = {{Early galaxy formation and its large-scale effects}},
  year    = {2018},
  month   = dec,
  pages   = {1-64},
  volume  = {780},
  adsurl  = {https://ui.adsabs.harvard.edu/abs/2018PhR...780....1D},
  doi     = {10.1016/j.physrep.2018.10.002},
  eprint  = {1809.09136},
}

@Article{ODonnell+1994,
  author  = {{O'Donnell}, James E.},
  journal = {\apj},
  title   = {{R v-dependent Optical and Near-Ultraviolet Extinction}},
  year    = {1994},
  month   = feb,
  pages   = {158},
  volume  = {422},
  adsurl  = {https://ui.adsabs.harvard.edu/abs/1994ApJ...422..158O},
  doi     = {10.1086/173713},
}

@Article{Bruzual+2003,
  author  = {{Bruzual}, G. and {Charlot}, S.},
  journal = {\mnras},
  title   = {{Stellar population synthesis at the resolution of 2003}},
  year    = {2003},
  month   = oct,
  number  = {4},
  pages   = {1000-1028},
  volume  = {344},
  adsurl  = {https://ui.adsabs.harvard.edu/abs/2003MNRAS.344.1000B},
  doi     = {10.1046/j.1365-8711.2003.06897.x},
  eprint  = {astro-ph/0309134},
}

@Article{Trebitsch+2017,
  author  = {{Trebitsch}, Maxime and {Blaizot}, J{\'e}r{\'e}my and {Rosdahl}, Joakim and {Devriendt}, Julien and {Slyz}, Adrianne},
  journal = {\mnras},
  title   = {{Fluctuating feedback-regulated escape fraction of ionizing radiation in low-mass, high-redshift galaxies}},
  year    = {2017},
  month   = sep,
  number  = {1},
  pages   = {224-239},
  volume  = {470},
  adsurl  = {https://ui.adsabs.harvard.edu/abs/2017MNRAS.470..224T},
  doi     = {10.1093/mnras/stx1060},
  eprint  = {1705.00941},
}

@Article{Kim+2019,
  author  = {{Kim}, Jeong-Gyu and {Kim}, Woong-Tae and {Ostriker}, Eve C.},
  journal = {\apj},
  title   = {{Modeling UV Radiation Feedback from Massive Stars. III. Escape of Radiation from Star-forming Giant Molecular Clouds}},
  year    = {2019},
  month   = sep,
  number  = {1},
  pages   = {102},
  volume  = {883},
  adsurl  = {https://ui.adsabs.harvard.edu/abs/2019ApJ...883..102K},
  doi     = {10.3847/1538-4357/ab3d3d},
  eid     = {102},
  eprint  = {1908.07549},
}

@Article{Howard+2017,
  author  = {{Howard}, Corey and {Pudritz}, Ralph and {Klessen}, Ralf},
  journal = {\apj},
  title   = {{Ultraviolet Escape Fractions from Giant Molecular Clouds during Early Cluster Formation}},
  year    = {2017},
  month   = jan,
  number  = {1},
  pages   = {40},
  volume  = {834},
  adsurl  = {https://ui.adsabs.harvard.edu/abs/2017ApJ...834...40H},
  doi     = {10.3847/1538-4357/834/1/40},
  eid     = {40},
  eprint  = {1611.02708},
}

@Article{Kakiichi+2021,
  author  = {{Kakiichi}, Koki and {Gronke}, Max},
  journal = {\apj},
  title   = {{Radiation Hydrodynamics of Turbulent H II Regions in Molecular Clouds: A Physical Origin of LyC Leakage and the Associated Ly{\ensuremath{\alpha}} Spectra}},
  year    = {2021},
  month   = feb,
  number  = {1},
  pages   = {30},
  volume  = {908},
  adsurl  = {https://ui.adsabs.harvard.edu/abs/2021ApJ...908...30K},
  doi     = {10.3847/1538-4357/abc2d9},
  eid     = {30},
  eprint  = {1905.02480},
}

@Article{Howard+2018,
  author  = {{Howard}, Corey S. and {Pudritz}, Ralph E. and {Harris}, William E. and {Klessen}, Ralf S.},
  journal = {\mnras},
  title   = {{Simulating the UV escape fractions from molecular cloud populations in star-forming dwarf and spiral galaxies}},
  year    = {2018},
  month   = apr,
  number  = {3},
  pages   = {3121-3134},
  volume  = {475},
  adsurl  = {https://ui.adsabs.harvard.edu/abs/2018MNRAS.475.3121H},
  doi     = {10.1093/mnras/stx3276},
  eprint  = {1710.04283},
}

@Article{Hummer+1987,
  author  = {{Hummer}, D.~G. and {Storey}, P.~J.},
  journal = {\mnras},
  title   = {{Recombination-line intensities for hydrogenic ions - I. Case B calculations for H I and He II.}},
  year    = {1987},
  month   = feb,
  pages   = {801-820},
  volume  = {224},
  adsurl  = {https://ui.adsabs.harvard.edu/abs/1987MNRAS.224..801H},
  doi     = {10.1093/mnras/224.3.801},
}

@Article{Astropy+2022,
  author  = {{Astropy Collaboration} and {Price-Whelan}, Adrian M. and {Lim}, Pey Lian and {Earl}, Nicholas and {Starkman}, Nathaniel and {Bradley}, Larry and {Shupe}, David L. and {Patil}, Aarya A. and {Corrales}, Lia and {Brasseur}, C.~E. and {N{\"o}the}, Maximilian and {Donath}, Axel and {Tollerud}, Erik and {Morris}, Brett M. and {Ginsburg}, Adam and {Vaher}, Eero and {Weaver}, Benjamin A. and {Tocknell}, James and {Jamieson}, William and {van Kerkwijk}, Marten H. and {Robitaille}, Thomas P. and {Merry}, Bruce and {Bachetti}, Matteo and {G{\"u}nther}, H. Moritz and {Aldcroft}, Thomas L. and {Alvarado-Montes}, Jaime A. and {Archibald}, Anne M. and {B{\'o}di}, Attila and {Bapat}, Shreyas and {Barentsen}, Geert and {Baz{\'a}n}, Juanjo and {Biswas}, Manish and {Boquien}, M{\'e}d{\'e}ric and {Burke}, D.~J. and {Cara}, Daria and {Cara}, Mihai and {Conroy}, Kyle E. and {Conseil}, Simon and {Craig}, Matthew W. and {Cross}, Robert M. and {Cruz}, Kelle L. and {D'Eugenio}, Francesco and {Dencheva}, Nadia and {Devillepoix}, Hadrien A.~R. and {Dietrich}, J{\"o}rg P. and {Eigenbrot}, Arthur Davis and {Erben}, Thomas and {Ferreira}, Leonardo and {Foreman-Mackey}, Daniel and {Fox}, Ryan and {Freij}, Nabil and {Garg}, Suyog and {Geda}, Robel and {Glattly}, Lauren and {Gondhalekar}, Yash and {Gordon}, Karl D. and {Grant}, David and {Greenfield}, Perry and {Groener}, Austen M. and {Guest}, Steve and {Gurovich}, Sebastian and {Handberg}, Rasmus and {Hart}, Akeem and {Hatfield-Dodds}, Zac and {Homeier}, Derek and {Hosseinzadeh}, Griffin and {Jenness}, Tim and {Jones}, Craig K. and {Joseph}, Prajwel and {Kalmbach}, J. Bryce and {Karamehmetoglu}, Emir and {Ka{\l}uszy{\'n}ski}, Miko{\l}aj and {Kelley}, Michael S.~P. and {Kern}, Nicholas and {Kerzendorf}, Wolfgang E. and {Koch}, Eric W. and {Kulumani}, Shankar and {Lee}, Antony and {Ly}, Chun and {Ma}, Zhiyuan and {MacBride}, Conor and {Maljaars}, Jakob M. and {Muna}, Demitri and {Murphy}, N.~A. and {Norman}, Henrik and {O'Steen}, Richard and {Oman}, Kyle A. and {Pacifici}, Camilla and {Pascual}, Sergio and {Pascual-Granado}, J. and {Patil}, Rohit R. and {Perren}, Gabriel I. and {Pickering}, Timothy E. and {Rastogi}, Tanuj and {Roulston}, Benjamin R. and {Ryan}, Daniel F. and {Rykoff}, Eli S. and {Sabater}, Jose and {Sakurikar}, Parikshit and {Salgado}, Jes{\'u}s and {Sanghi}, Aniket and {Saunders}, Nicholas and {Savchenko}, Volodymyr and {Schwardt}, Ludwig and {Seifert-Eckert}, Michael and {Shih}, Albert Y. and {Jain}, Anany Shrey and {Shukla}, Gyanendra and {Sick}, Jonathan and {Simpson}, Chris and {Singanamalla}, Sudheesh and {Singer}, Leo P. and {Singhal}, Jaladh and {Sinha}, Manodeep and {Sip{\H{o}}cz}, Brigitta M. and {Spitler}, Lee R. and {Stansby}, David and {Streicher}, Ole and {{\v{S}}umak}, Jani and {Swinbank}, John D. and {Taranu}, Dan S. and {Tewary}, Nikita and {Tremblay}, Grant R. and {Val-Borro}, Miguel de and {Van Kooten}, Samuel J. and {Vasovi{\'c}}, Zlatan and {Verma}, Shresth and {de Miranda Cardoso}, Jos{\'e} Vin{\'\i}cius and {Williams}, Peter K.~G. and {Wilson}, Tom J. and {Winkel}, Benjamin and {Wood-Vasey}, W.~M. and {Xue}, Rui and {Yoachim}, Peter and {Zhang}, Chen and {Zonca}, Andrea and {Astropy Project Contributors}},
  journal = {\apj},
  title   = {{The Astropy Project: Sustaining and Growing a Community-oriented Open-source Project and the Latest Major Release (v5.0) of the Core Package}},
  year    = {2022},
  month   = aug,
  number  = {2},
  pages   = {167},
  volume  = {935},
  adsurl  = {https://ui.adsabs.harvard.edu/abs/2022ApJ...935..167A},
  doi     = {10.3847/1538-4357/ac7c74},
  eid     = {167},
  eprint  = {2206.14220},
}

@Article{Schlafly+2011,
  author  = {{Schlafly}, Edward F. and {Finkbeiner}, Douglas P.},
  journal = {\apj},
  title   = {{Measuring Reddening with Sloan Digital Sky Survey Stellar Spectra and Recalibrating SFD}},
  year    = {2011},
  month   = aug,
  number  = {2},
  pages   = {103},
  volume  = {737},
  adsurl  = {https://ui.adsabs.harvard.edu/abs/2011ApJ...737..103S},
  doi     = {10.1088/0004-637X/737/2/103},
  eid     = {103},
  eprint  = {1012.4804},
}

@Article{Doran+2013,
  author  = {{Doran}, E.~I. and {Crowther}, P.~A. and {de Koter}, A. and {Evans}, C.~J. and {McEvoy}, C. and {Walborn}, N.~R. and {Bastian}, N. and {Bestenlehner}, J.~M. and {Gr{\"a}fener}, G. and {Herrero}, A. and {K{\"o}hler}, K. and {Ma{\'\i}z Apell{\'a}niz}, J. and {Najarro}, F. and {Puls}, J. and {Sana}, H. and {Schneider}, F.~R.~N. and {Taylor}, W.~D. and {van Loon}, J. Th. and {Vink}, J.~S.},
  journal = {\aap},
  title   = {{The VLT-FLAMES Tarantula Survey. XI. A census of the hot luminous stars and their feedback in 30 Doradus}},
  year    = {2013},
  month   = oct,
  pages   = {A134},
  volume  = {558},
  adsurl  = {https://ui.adsabs.harvard.edu/abs/2013A&A...558A.134D},
  doi     = {10.1051/0004-6361/201321824},
  eid     = {A134},
  eprint  = {1308.3412},
}

@Article{Grasha+2015,
  author  = {{Grasha}, K. and {Calzetti}, D. and {Adamo}, A. and {Kim}, H. and {Elmegreen}, B.~G. and {Gouliermis}, D.~A. and {Aloisi}, A. and {Bright}, S.~N. and {Christian}, C. and {Cignoni}, M. and {Dale}, D.~A. and {Dobbs}, C. and {Elmegreen}, D.~M. and {Fumagalli}, M. and {Gallagher}, J.~S., III and {Grebel}, E.~K. and {Johnson}, K.~E. and {Lee}, J.~C. and {Messa}, M. and {Smith}, L.~J. and {Ryon}, J.~E. and {Thilker}, D. and {Ubeda}, L. and {Wofford}, A.},
  journal = {\apj},
  title   = {{The Spatial Distribution of the Young Stellar Clusters in the Star-forming Galaxy NGC 628}},
  year    = {2015},
  month   = dec,
  number  = {2},
  pages   = {93},
  volume  = {815},
  adsurl  = {https://ui.adsabs.harvard.edu/abs/2015ApJ...815...93G},
  doi     = {10.1088/0004-637X/815/2/93},
  eid     = {93},
  eprint  = {1511.02233},
}

@Article{Wei+2020,
  author  = {{Wei}, Wei and {Huerta}, E.~A. and {Whitmore}, Bradley C. and {Lee}, Janice C. and {Hannon}, Stephen and {Chandar}, Rupali and {Dale}, Daniel A. and {Larson}, Kirsten L. and {Thilker}, David A. and {Ubeda}, Leonardo and {Boquien}, M{\'e}d{\'e}ric and {Chevance}, M{\'e}lanie and {Kruijssen}, J.~M. Diederik and {Schruba}, Andreas and {Blanc}, Guillermo A. and {Congiu}, Enrico},
  journal = {\mnras},
  title   = {{Deep transfer learning for star cluster classification: I. application to the PHANGS-HST survey}},
  year    = {2020},
  month   = apr,
  number  = {3},
  pages   = {3178-3193},
  volume  = {493},
  adsurl  = {https://ui.adsabs.harvard.edu/abs/2020MNRAS.493.3178W},
  doi     = {10.1093/mnras/staa325},
  eprint  = {1909.02024},
}

@Article{Georgy+2013,
  author  = {{Georgy}, C. and {Ekstr{\"o}m}, S. and {Eggenberger}, P. and {Meynet}, G. and {Haemmerl{\'e}}, L. and {Maeder}, A. and {Granada}, A. and {Groh}, J.~H. and {Hirschi}, R. and {Mowlavi}, N. and {Yusof}, N. and {Charbonnel}, C. and {Decressin}, T. and {Barblan}, F.},
  journal = {\aap},
  title   = {{Grids of stellar models with rotation. III. Models from 0.8 to 120 M$_\odot$ at a metallicity Z = 0.002}},
  year    = {2013},
  month   = Oct,
  pages   = {A103},
  volume  = {558},
  adsurl  = {https://ui.adsabs.harvard.edu/#abs/2013A&A...558A.103G},
  doi     = {10.1051/0004-6361/201322178},
  eid     = {A103},
  eprint  = {1308.2914},
}

@Article{Hassani+2024,
  author  = {{Hassani}, Hamid and {Rosolowsky}, Erik and {Koch}, Eric W. and {Postma}, Joseph and {Nofech}, Joseph and {Corbould}, Harrisen and {Thilker}, David and {Leroy}, Adam K. and {Schinnerer}, Eva and {Belfiore}, Francesco and {Bigiel}, Frank and {Boquien}, M{\'e}d{\'e}ric and {Chevance}, M{\'e}lanie and {Dale}, Daniel A. and {Egorov}, Oleg V. and {Emsellem}, Eric and {Glover}, Simon C.~O. and {Grasha}, Kathryn and {Groves}, Brent and {Henny}, Kiana and {Kim}, Jaeyeon and {Klessen}, Ralf S. and {Kreckel}, Kathryn and {Kruijssen}, J.~M. Diederik and {Lee}, Janice C. and {Lopez}, Laura A. and {Neumann}, Justus and {Pan}, Hsi-An and {Sandstrom}, Karin M. and {Sarbadhicary}, Sumit K. and {Sun}, Jiayi and {Williams}, Thomas G.},
  journal = {\apjs},
  title   = {{The PHANGS-AstroSat Atlas of Nearby Star-forming Galaxies}},
  year    = {2024},
  month   = mar,
  number  = {1},
  pages   = {2},
  volume  = {271},
  adsurl  = {https://ui.adsabs.harvard.edu/abs/2024ApJS..271....2H},
  doi     = {10.3847/1538-4365/ad152c},
  eid     = {2},
  eprint  = {2312.06031},
}

@Article{Pellegrini+2012,
  author  = {{Pellegrini}, E.~W. and {Oey}, M.~S. and {Winkler}, P.~F. and {Points}, S.~D. and {Smith}, R.~C. and {Jaskot}, A.~E. and {Zastrow}, J.},
  journal = {\apj},
  title   = {{The Optical Depth of H II Regions in the Magellanic Clouds}},
  year    = {2012},
  month   = aug,
  number  = {1},
  pages   = {40},
  volume  = {755},
  adsurl  = {https://ui.adsabs.harvard.edu/abs/2012ApJ...755...40P},
  doi     = {10.1088/0004-637X/755/1/40},
  eid     = {40},
  eprint  = {1202.3334},
}

@Article{Teh+2023,
  author  = {{Teh}, Jia Wei and {Grasha}, Kathryn and {Krumholz}, Mark R. and {Battisti}, Andrew J. and {Calzetti}, Daniela and {Rousseau-Nepton}, Laurie and {Rhea}, Carter and {Adamo}, Angela and {Kennicutt}, Robert C. and {Grebel}, Eva K. and {Cook}, David O. and {Combes}, Francoise and {Messa}, Matteo and {Linden}, Sean T. and {Klessen}, Ralf S. and {Vilchez}, Jos{\'e} M. and {Fumagalli}, Michele and {McLeod}, Anna and {Smith}, Linda J. and {Chemin}, Laurent and {Wang}, Junfeng and {Sabbi}, Elena and {Sacchi}, Elena and {Petric}, Andreea and {Della Bruna}, Lorenza and {Boselli}, Alessandro},
  journal = {\mnras},
  title   = {{Constraining the LyC escape fraction from LEGUS star clusters with SIGNALS H II region observations: a pilot study of NGC 628}},
  year    = {2023},
  month   = sep,
  number  = {1},
  pages   = {1191-1210},
  volume  = {524},
  adsurl  = {https://ui.adsabs.harvard.edu/abs/2023MNRAS.524.1191T},
  doi     = {10.1093/mnras/stad1780},
  eprint  = {2306.05457},
}

@Article{Rahner+2017,
  author  = {{Rahner}, D. and {Pellegrini}, E.~W. and {Glover}, S.~C.~O. and {Klessen}, R.~S.},
  journal = {\mnras},
  title   = {{Winds and radiation in unison: a new semi-analytic feedback model for cloud dissolution}},
  year    = {2017},
  month   = oct,
  pages   = {4453-4472},
  volume  = {470},
  adsurl  = {http://adsabs.harvard.edu/abs/2017MNRAS.470.4453R},
  doi     = {10.1093/mnras/stx1532},
  eprint  = {1704.04240},
}

@Article{Scheuermann+2023,
  author  = {{Scheuermann}, Fabian and {Kreckel}, Kathryn and {Barnes}, Ashley T. and {Belfiore}, Francesco and {Groves}, Brent and {Hannon}, Stephen and {Lee}, Janice C. and {Minsley}, Rebecca and {Rosolowsky}, Erik and {Bigiel}, Frank and {Blanc}, Guillermo A. and {Boquien}, M{\'e}d{\'e}ric and {Dale}, Daniel A. and {Deger}, Sinan and {Egorov}, Oleg V. and {Emsellem}, Eric and {Glover}, Simon C.~O. and {Grasha}, Kathryn and {Hassani}, Hamid and {Jeffreson}, Sarah M.~R. and {Klessen}, Ralf S. and {Kruijssen}, J.~M. Diederik and {Larson}, Kirsten L. and {Leroy}, Adam K. and {Lopez}, Laura A. and {Pan}, Hsi-An and {S{\'a}nchez-Bl{\'a}zquez}, Patricia and {Santoro}, Francesco and {Schinnerer}, Eva and {Thilker}, David A. and {Whitmore}, Bradley C. and {Watkins}, Elizabeth J. and {Williams}, Thomas G.},
  journal = {\mnras},
  title   = {{Stellar associations powering H II regions - I. Defining an evolutionary sequence}},
  year    = {2023},
  month   = jun,
  number  = {2},
  pages   = {2369-2383},
  volume  = {522},
  adsurl  = {https://ui.adsabs.harvard.edu/abs/2023MNRAS.522.2369S},
  doi     = {10.1093/mnras/stad878},
  eprint  = {2303.12101},
}

@Article{Georgy+2012,
  author  = {{Georgy}, C. and {Ekstr{\"o}m}, S. and {Meynet}, G. and {Massey}, P. and {Levesque}, E.~M. and {Hirschi}, R. and {Eggenberger}, P. and {Maeder}, A.},
  journal = {\aap},
  title   = {{Grids of stellar models with rotation. II. WR populations and supernovae/GRB progenitors at Z = 0.014}},
  year    = {2012},
  month   = Jun,
  pages   = {A29},
  volume  = {542},
  adsurl  = {https://ui.adsabs.harvard.edu/#abs/2012A&A...542A..29G},
  doi     = {10.1051/0004-6361/201118340},
  eid     = {A29},
  eprint  = {1203.5243},
}

@Article{Leitherer+1999,
  author  = {{Leitherer}, C. and {Schaerer}, D. and {Goldader}, J.~D. and {Delgado}, R.~M.~G. and {Robert}, C. and {Kune}, D.~F. and {de Mello}, D.~F. and {Devost}, D. and {Heckman}, T.~M.},
  journal = {\apjs},
  title   = {{Starburst99: Synthesis Models for Galaxies with Active Star Formation}},
  year    = {1999},
  month   = jul,
  pages   = {3-40},
  volume  = {123},
  adsurl  = {http://adsabs.harvard.edu/abs/1999ApJS..123....3L},
  doi     = {10.1086/313233},
  eprint  = {astro-ph/9902334},
}

@Article{Chabrier+2003,
  author  = {{Chabrier}, Gilles},
  journal = {\pasp},
  title   = {{Galactic Stellar and Substellar Initial Mass Function}},
  year    = {2003},
  month   = jul,
  number  = {809},
  pages   = {763-795},
  volume  = {115},
  adsurl  = {https://ui.adsabs.harvard.edu/abs/2003PASP..115..763C},
  doi     = {10.1086/376392},
  eprint  = {astro-ph/0304382},
}

@Article{Klessen+2016,
  author    = {{Klessen}, Ralf S. and {Glover}, Simon C.~O.},
  journal   = {Saas-Fee Advanced Course},
  title     = {{Physical Processes in the Interstellar Medium}},
  year      = {2016},
  month     = jan,
  pages     = {85--249},
  volume    = {43},
  adsurl    = {https://ui.adsabs.harvard.edu/abs/2016SAAS...43...85K},
  booktitle = {Star Formation in Galaxy Evolution: Connecting Numerical Models to Reality},
  doi       = {10.1007/978-3-662-47890-5_2},
  eprint    = {1412.5182},
  publisher = {Springer Berlin Heidelberg},
}

@Article{Ekstroem+2012,
  author  = {{Ekstr{\"o}m}, S. and {Georgy}, C. and {Eggenberger}, P. and {Meynet}, G. and {Mowlavi}, N. and {Wyttenbach}, A. and {Granada}, A. and {Decressin}, T. and {Hirschi}, R. and {Frischknecht}, U. and {Charbonnel}, C. and {Maeder}, A.},
  journal = {\aap},
  title   = {{Grids of stellar models with rotation. I. Models from 0.8 to 120 M$_\odot$ at solar metallicity (Z = 0.014)}},
  year    = {2012},
  month   = jan,
  pages   = {A146},
  volume  = {537},
  adsurl  = {http://adsabs.harvard.edu/abs/2012A%26A...537A.146E},
  doi     = {10.1051/0004-6361/201117751},
  eid     = {A146},
  eprint  = {1110.5049},
}

@Article{Eldridge+2017,
  author  = {{Eldridge}, J.~J. and {Stanway}, E.~R. and {Xiao}, L. and {McClelland}, L.~A.~S. and {Taylor}, G. and {Ng}, M. and {Greis}, S.~M.~L. and {Bray}, J.~C.},
  journal = {\pasa},
  title   = {{Binary Population and Spectral Synthesis Version 2.1: Construction, Observational Verification, and New Results}},
  year    = {2017},
  month   = nov,
  pages   = {e058},
  volume  = {34},
  adsurl  = {https://ui.adsabs.harvard.edu/abs/2017PASA...34...58E},
  doi     = {10.1017/pasa.2017.51},
  eid     = {e058},
  eprint  = {1710.02154},
}

@Article{Astropy+2013,
  author  = {{Astropy Collaboration} and {Robitaille}, Thomas P. and {Tollerud}, Erik J. and {Greenfield}, Perry and {Droettboom}, Michael and {Bray}, Erik and {Aldcroft}, Tom and {Davis}, Matt and {Ginsburg}, Adam and {Price-Whelan}, Adrian M. and {Kerzendorf}, Wolfgang E. and {Conley}, Alexander and {Crighton}, Neil and {Barbary}, Kyle and {Muna}, Demitri and {Ferguson}, Henry and {Grollier}, Fr{\'e}d{\'e}ric and {Parikh}, Madhura M. and {Nair}, Prasanth H. and {Unther}, Hans M. and {Deil}, Christoph and {Woillez}, Julien and {Conseil}, Simon and {Kramer}, Roban and {Turner}, James E.~H. and {Singer}, Leo and {Fox}, Ryan and {Weaver}, Benjamin A. and {Zabalza}, Victor and {Edwards}, Zachary I. and {Azalee Bostroem}, K. and {Burke}, D.~J. and {Casey}, Andrew R. and {Crawford}, Steven M. and {Dencheva}, Nadia and {Ely}, Justin and {Jenness}, Tim and {Labrie}, Kathleen and {Lim}, Pey Lian and {Pierfederici}, Francesco and {Pontzen}, Andrew and {Ptak}, Andy and {Refsdal}, Brian and {Servillat}, Mathieu and {Streicher}, Ole},
  journal = {\aap},
  title   = {{Astropy: A community Python package for astronomy}},
  year    = {2013},
  month   = oct,
  pages   = {A33},
  volume  = {558},
  adsurl  = {https://ui.adsabs.harvard.edu/abs/2013A&A...558A..33A},
  doi     = {10.1051/0004-6361/201322068},
  eid     = {A33},
  eprint  = {1307.6212},
}

@Article{Paardekooper+2011,
  author  = {{Paardekooper}, J. -P. and {Pelupessy}, F.~I. and {Altay}, G. and {Kruip}, C.~J.~H.},
  journal = {\aap},
  title   = {{The escape of ionising radiation from high-redshift dwarf galaxies}},
  year    = {2011},
  month   = jun,
  pages   = {A87},
  volume  = {530},
  adsurl  = {https://ui.adsabs.harvard.edu/abs/2011A&A...530A..87P},
  doi     = {10.1051/0004-6361/201116841},
  eid     = {A87},
  eprint  = {1104.3584},
}

@Article{Astropy+2018,
  author  = {{Astropy Collaboration} and {Price-Whelan}, A.~M. and {Sip{\H{o}}cz}, B.~M. and {G{\"u}nther}, H.~M. and {Lim}, P.~L. and {Crawford}, S.~M. and {Conseil}, S. and {Shupe}, D.~L. and {Craig}, M.~W. and {Dencheva}, N. and {Ginsburg}, A. and {Vand erPlas}, J.~T. and {Bradley}, L.~D. and {P{\'e}rez-Su{\'a}rez}, D. and {de Val-Borro}, M. and {Aldcroft}, T.~L. and {Cruz}, K.~L. and {Robitaille}, T.~P. and {Tollerud}, E.~J. and {Ardelean}, C. and {Babej}, T. and {Bach}, Y.~P. and {Bachetti}, M. and {Bakanov}, A.~V. and {Bamford}, S.~P. and {Barentsen}, G. and {Barmby}, P. and {Baumbach}, A. and {Berry}, K.~L. and {Biscani}, F. and {Boquien}, M. and {Bostroem}, K.~A. and {Bouma}, L.~G. and {Brammer}, G.~B. and {Bray}, E.~M. and {Breytenbach}, H. and {Buddelmeijer}, H. and {Burke}, D.~J. and {Calderone}, G. and {Cano Rodr{\'\i}guez}, J.~L. and {Cara}, M. and {Cardoso}, J.~V.~M. and {Cheedella}, S. and {Copin}, Y. and {Corrales}, L. and {Crichton}, D. and {D'Avella}, D. and {Deil}, C. and {Depagne}, {\'E}. and {Dietrich}, J.~P. and {Donath}, A. and {Droettboom}, M. and {Earl}, N. and {Erben}, T. and {Fabbro}, S. and {Ferreira}, L.~A. and {Finethy}, T. and {Fox}, R.~T. and {Garrison}, L.~H. and {Gibbons}, S.~L.~J. and {Goldstein}, D.~A. and {Gommers}, R. and {Greco}, J.~P. and {Greenfield}, P. and {Groener}, A.~M. and {Grollier}, F. and {Hagen}, A. and {Hirst}, P. and {Homeier}, D. and {Horton}, A.~J. and {Hosseinzadeh}, G. and {Hu}, L. and {Hunkeler}, J.~S. and {Ivezi{\'c}}, {\v{Z}}. and {Jain}, A. and {Jenness}, T. and {Kanarek}, G. and {Kendrew}, S. and {Kern}, N.~S. and {Kerzendorf}, W.~E. and {Khvalko}, A. and {King}, J. and {Kirkby}, D. and {Kulkarni}, A.~M. and {Kumar}, A. and {Lee}, A. and {Lenz}, D. and {Littlefair}, S.~P. and {Ma}, Z. and {Macleod}, D.~M. and {Mastropietro}, M. and {McCully}, C. and {Montagnac}, S. and {Morris}, B.~M. and {Mueller}, M. and {Mumford}, S.~J. and {Muna}, D. and {Murphy}, N.~A. and {Nelson}, S. and {Nguyen}, G.~H. and {Ninan}, J.~P. and {N{\"o}the}, M. and {Ogaz}, S. and {Oh}, S. and {Parejko}, J.~K. and {Parley}, N. and {Pascual}, S. and {Patil}, R. and {Patil}, A.~A. and {Plunkett}, A.~L. and {Prochaska}, J.~X. and {Rastogi}, T. and {Reddy Janga}, V. and {Sabater}, J. and {Sakurikar}, P. and {Seifert}, M. and {Sherbert}, L.~E. and {Sherwood-Taylor}, H. and {Shih}, A.~Y. and {Sick}, J. and {Silbiger}, M.~T. and {Singanamalla}, S. and {Singer}, L.~P. and {Sladen}, P.~H. and {Sooley}, K.~A. and {Sornarajah}, S. and {Streicher}, O. and {Teuben}, P. and {Thomas}, S.~W. and {Tremblay}, G.~R. and {Turner}, J.~E.~H. and {Terr{\'o}n}, V. and {van Kerkwijk}, M.~H. and {de la Vega}, A. and {Watkins}, L.~L. and {Weaver}, B.~A. and {Whitmore}, J.~B. and {Woillez}, J. and {Zabalza}, V. and {Astropy Contributors}},
  journal = {\aj},
  title   = {{The Astropy Project: Building an Open-science Project and Status of the v2.0 Core Package}},
  year    = {2018},
  month   = sep,
  number  = {3},
  pages   = {123},
  volume  = {156},
  adsurl  = {https://ui.adsabs.harvard.edu/abs/2018AJ....156..123A},
  doi     = {10.3847/1538-3881/aabc4f},
  eid     = {123},
  eprint  = {1801.02634},
}

\clearpage
\begin{appendix}

\section{Ionizing photon flux from different models}\label{sec:appendix_models}

In \cref{fig:ionizing_photon_flux_models} we compare the ionizing photon flux $Q(\mathrm{H}^0)$ for different models (top panel) and metallicities (lower panel). 
A detailed listing of the models can be found in \cref{tbl:stellar_models}. 
These plots demonstrate that, at constant metallicity, all models except the one with rotation are similar for the first few \si{\mega\year}. 
Around \SI{6}{\mega\year}, however, the effects of binarity becomes relevant as its flux decreases much more slowly than the other models. 
For the impact of metallicity, we include the low-metallicity models from \textsc{starburst99}, based on the stellar model by \citet{Georgy+2013}. 
The variation between models that are close to solar metallicity is practically negligible, while the predicted flux is slightly higher at lower metallicities. 
Given that the flux decreases by an order of magnitude every few million years and the uncertainty of the age is comparatively high, the choice of model is likely to have only a minor influence. 

\begin{figure}
    \centering
    \includegraphics[width=\columnwidth]{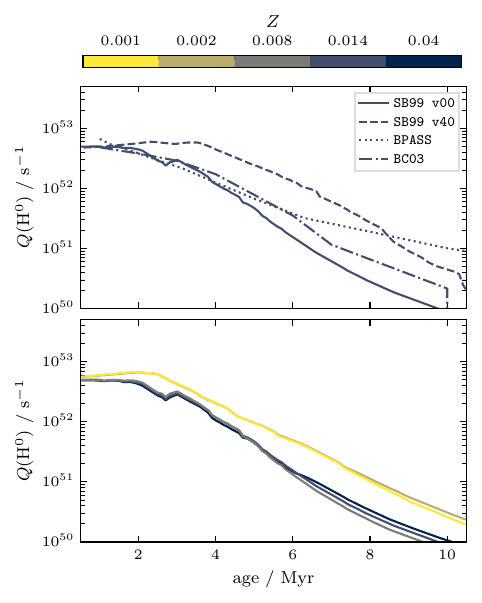}
    \caption[Ionizing photon flux $\Qpred$ as a function of age from different models and metallicities.]{Ionizing photon flux $\Qpred$ of a star cluster with a mass of \SI{e6}{\Msun} as a function of age. 
    The top panel shows the values predicted by different models, all at solar metallicity, while the lower panel shows the \texttt{SB99 v00} model at various metallicities.} 
    \label{fig:ionizing_photon_flux_models}
\end{figure}

\begin{table*}
    \centering
    \caption{Codes used to compute the ionizing photon flux $Q(\mathrm{H}^0)$.}
    \begin{tabular}{llll}
        \toprule
        Name & Code & Stellar model & IMF \\\midrule
        \texttt{BC03} & \textsc{cigale} & BC03 & \\
        & \citep{Boquien+2019} & \citep{Bruzual+2003} & \citep{Chabrier+2003} \\\addlinespace
        \texttt{SB99 v00} & \textsc{starburst99} & Geneva without rotation & \\
        & \citep{Leitherer+2014} & \citep{Ekstroem+2012} & \citep{Kroupa+2001} \\\addlinespace
        \texttt{SB99 v40} & \textsc{starburst99} & Geneva with rotation & \\
        & \citep{Leitherer+2014} & \citep{Ekstroem+2012} & \citep{Kroupa+2001} \\\addlinespace
        \texttt{BPASS} & \textsc{bpass} & & \\
        & \citep{Eldridge+2009} & & \citep{Kroupa+2001} \\\addlinespace
        \texttt{SLUG} & \textsc{slug} & Geneva & \\
        & \citep{daSilva+2012} & \citep{Ekstroem+2012} & \citep{Kroupa+2001} \\\bottomrule
    \end{tabular}
    \label{tbl:stellar_models}
\end{table*}

\section{Comparison with Teh et al.\ (2023)}\label{sec:appendix_slug}

\begin{figure}
    \centering
    \includegraphics[width=0.75\columnwidth]{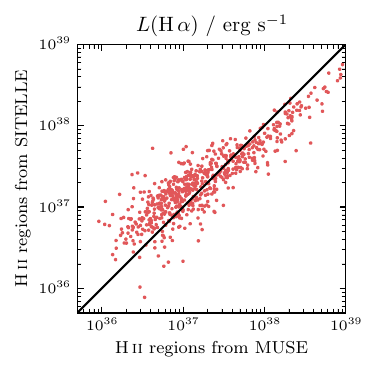}
    \caption{Comparisons between the \HII regions from PHANGS \citep[][observed with MUSE]{Groves+2023} and the ones from SIGNALS \citep[][observed with SITELLE]{RousseauNepton+2019}.}
    \label{fig:HII_region_Teh}
\end{figure}

The stellar catalogues presented in \cref{sec:data:stars} rely on the assumption that the clusters and associations in our sample are massive enough such that a deterministic relationship between the physical and photometric properties of the stellar populations exists. 
While we try to ensure this by focusing our analysis on high mass objects from the \texttt{robust} subsamples defined in \cref{sec:matched_catalogues}, this means that we are omitting a large part of the sample. 
Another approach is to employ a stochastic stellar population synthesis code like \textsc{slug} \citep{daSilva+2012, daSilva+2014} to infer a distribution of possible values for $\Qpred$. 

A particularly relevant literature result is the study by \citet{Teh+2023}, who measured the escape fraction of \galaxyname{NGC}{0628} (which is also in our sample) with the help of \textsc{slug}. 
They combined \HII regions from the Star formation, ionized gas and nebular abundances legacy (\mbox{SIGNALS}) survey \citep{RousseauNepton+2018} with star clusters from \mbox{LEGUS} \citep{Grasha+2015,Adamo+2017}. 
A large fraction of the 139 objects in their matched catalogue have negative escape fractions and based on a subsample of 42 objects, they measured a comparatively low value of $\fesc=\SI{9\pm6}{\percent}$. 
The striking difference to our results asks for a closer comparison. 

We start by comparing the \HII regions from MUSE to those from SITELLE and \num{643} of them match within $\SI{0.8}{\arcsec}$. 
In \cref{fig:HII_region_Teh} we compare the measured fluxes, and while the ones identified by SIGNALS are more luminous at the fainter end, the ones from PHANGS are brighter for the most luminous objects. 

To compare the stellar component, we take the five-band filters from our association catalogue and redo the SED fit with \textsc{slug}, following the procedure described in \citet{Teh+2023}:
First, a library of $\num{e7}$ stellar populations is generated to account for the effects of stochasticity in the IMF. 
For this, we adopt the following physical parameters: Solar metallicity, the IMF by \citet{Kroupa+2001}, Geneva evolutionary tracks \citep{Ekstroem+2012,Georgy+2012}, \textsc{starburst99} stellar atmosphere models \citep{Leitherer+1999,Vazquez+2005}, and a Milky Way extinction curve \citep{Cardelli+1989}. 
Then we use a modified version of \texttt{cluster\_slug} \citep{DellaBruna+2021,DellaBruna+2022b,Teh+2023} to compute the posterior probability density function (PDF) of the age, mass, and ionizing luminosity \citep[see][for more details]{Krumholz+2015}. 
We represent the $\Qpred$ values with the median (50th percentile) and the $1\sigma$ uncertainty with the 16th--84th percentile of the PDF distribution. 

The result is shown in \cref{fig:SLUG_vs_CIGALE}. 
As far as the masses are concerned, there is a large scatter, but not systematic differences. 
There are significant deviations in the derived ages, with those from \textsc{slug} generally being older. 
This is particularly problematic in the case of very young stars. 
While the lower limit of the age distribution is often consistent with being \SI{3}{\mega\year} or younger, the median rarely is. 
Since the majority of the ionizing photons are emitted during this period, the predicted fluxes are generally lower.

\begin{figure*}
    \centering
    \includegraphics[width=\textwidth]{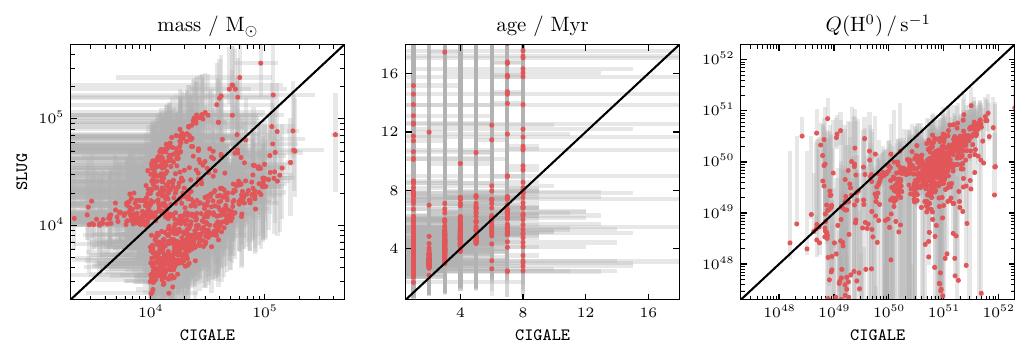}
    \caption[Properties of our association catalogue with SLUG.]{We ran \textsc{slug} on the stellar associations in the matched catalogue. 
    For the comparison we define a \texttt{robust} subsample, where either \textsc{cigale} or \textsc{slug} predict an age younger than $\SI{8}{\mega\year}$ and a mass above $\SI{e4}{\Msun}$ (859 objects). 
    The masses show a wide scatter in both directions, but \textsc{slug} generally predicts older ages. 
    As a result, the ionizing photon flux is systematically lower than the one from \textsc{cigale}.}
    \label{fig:SLUG_vs_CIGALE}
\end{figure*}

When matched to the \HII region catalogue from MUSE, we can compute the resulting escape fractions as shown in \cref{fig:fesc_histogram_slug}. 
Only 776 of them ($\sim\SI{19}{\percent}$) have positive escape fractions, and their escape fractions are uniformly distributed across the entire range from $\SIrange{0}{100}{\percent}$. 
If we focus on the young, massive, and contained objects that form a robust subsample, we find 228 objects that fulfil our criteria and only 50 of them ($\sim\SI{22}{\percent}$) have positive escape fractions. 
It was to be expected that \textsc{slug} would deliver different values at the faint end, due to stochastic sampling. 
Overall, it predicts systematically lower $\Qpred$, even at the bright end, where stochastic sampling should not be an issue. 
The lower $\Qpred$ is responsible for significantly lower escape fractions, which are mostly uniformly distributed between $\SIrange{0}{45}{\percent}$, and are in better agreement with results found by previous studies that are based on \textsc{slug} \citep{DellaBruna+2022b, Teh+2023}.

\begin{figure}
    \centering
    \includegraphics[width=0.75\columnwidth]{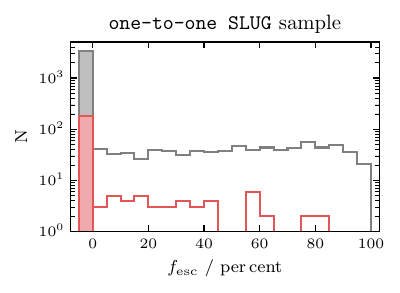}
    \caption[Histogram of the escape fractions based on stellar parameters inferred from SLUG.]{Histograms of the escape fractions based on stellar parameters inferred from \textsc{slug}. 
    The full sample is shown in gray and the robust subsample of young and massive, fully contained associations is shown in red. 
    All regions with negative escape fractions are grouped together in the shaded left bar.} 
    \label{fig:fesc_histogram_slug}
\end{figure}

\section{Completeness issues}\label{sec:appendix_completeness}

In \cref{sec:ionization_budget_galaxies} we consider the entire ionizing photon budget of each galaxy. 
In this scenario, different completeness limits for the \HII regions and stellar populations could bias those results. 
We therefore estimate the completeness for both to assess the contribution of missed objects. 
In \cref{fig:cumulative_Q} we show the normalised cumulative ionizing photon flux to estimate the contribution of faint \HII regions and low-mass associations. 

In the case of \galaxyname{IC}{5332}, \galaxyname{NGC}{3351} and \galaxyname{NGC}{5068}, faint \HII regions (below the completeness limit from \citealt{Santoro+2022}) contribute more than $\SI{10}{\percent}$ of the total ionizing photon flux. 
The potentially missing $\HA$ flux could increase the escape fraction, but while the values of the first two galaxies are above the average (see \cref{tbl:ionizing_photons_galaxies}), both also appear on the stellar side, weakening this conclusion. 

For \galaxyname{IC}{5332}, \galaxyname{NGC}{1300}, \galaxyname{NGC}{1433}, \galaxyname{NGC}{1512}, \galaxyname{NGC}{2835} and \galaxyname{NGC}{3351}, the contribution of low-mass associations ($<\SI{5000}{\Msun}$) is larger than $\SI{10}{\percent}$. 
In contrast to the case of the \HII regions, the potentially overlooked clusters would lead to a decrease of the escape fraction. 
While half of galaxies have lower escape fractions, the other half, including the two galaxies mentioned above, have higher values. 

For the 12 other galaxies, the completeness limit should not be an issue, and even for the seven discussed above, there is no noticeable impact. 

\begin{figure}
    \centering
    \includegraphics[width=\columnwidth]{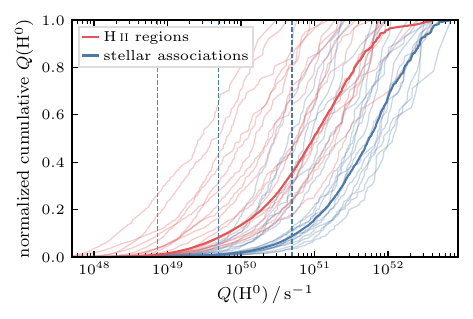}
    \caption[Normalized cumulative $Q(\mathrm{H}^0)$ as a function of $Q(\mathrm{H}^0)$ for \HII regions and stellar associations.]{Normalized cumulative $Q(\mathrm{H}^0)$ as a function of $Q(\mathrm{H}^0)$ for \HII regions and stellar associations. 
    The individual galaxies are shown in pastel and the sum of all galaxies is shown in bold colours. 
    We only show the \HII regions and associations in the overlapping coverage that do not fall in the centre (corresponding to the sample used in \cref{sec:ionization_budget_galaxies}). 
    The \HII region completeness limit from \citet{Santoro+2022}, corresponding to a luminosity of $\SI{e37}{\erg\per\second}$ is indicated by the left red dashed line. 
    The two blue lines corresponds to a $\SI{e3}{\Msun}$ and $\SI{e4}{\Msun}$ cluster that is $\SI{1}{\mega\year}$ old. 
    } 
    \label{fig:cumulative_Q}
\end{figure}

\end{appendix}

\end{document}